\documentclass{emulateapj}
\usepackage{epsfig,natbib}
\usepackage{graphicx}
\newcommand{\ms}{\mbox{m\,s$^{-1}~$}}
\newcommand{\kms}{\mbox{km\,s$^{-1}~$}}

\newcommand{\kse}{\mbox{km\,s$^{-1}$}}
\newcommand{\mse}{\mbox{m\,s$^{-1}$}}

\newcommand{\msun}{M$_{\odot}~$}

\newcommand{\rsun}{R$_{\odot}~$}

\newcommand{\lsun}{L$_{\odot}~$}

\newcommand{\mstar}{\ensuremath{M_{\star}}}

\newcommand{\chisq}{$\chi_{\nu}^2$}

\newcommand{\feh}{\ensuremath{[\mbox{Fe}/\mbox{H}]}}

\newcommand{\rphk}{\ensuremath{R'_{\mbox{\scriptsize HK}}}}

\newcommand{\lrphk}{\ensuremath{\log{\rphk}}}

\newcommand{\msini}{\ensuremath{M \sin i}}

\newcommand{\teff}{\ensuremath{T_{\rm eff}}}

\newcommand{\logg}{\ensuremath{\log{g}}}
\newcommand{\vsini}{\ensuremath{v \sin{i}}}

\newcommand{\bjdtdb}{\ensuremath{\rm {BJD_{TDB}}}}
\newcommand{\ecosw}{\ensuremath{e\cos{\omega_*}}}
\newcommand{\esinw}{\ensuremath{e\sin{\omega_*}}}
\newcommand{\fave}{\langle F \rangle}
\newcommand{\fluxcgs}{10$^9$ erg s$^{-1}$ cm$^{-2}$}

\newcommand{\mj}{\ensuremath{\,M_{\rm J}}}
\newcommand{\rj}{\ensuremath{\,R_{\rm J}}}
\newcommand{\Rp}{\ensuremath{1.86_{-0.16}^{+0.18}}}
\newcommand{\Mp}{\ensuremath{0.867_{-0.061}^{+0.065}}}

\shortauthors{Fulton {et~al.}}
\shorttitle{KELT-8b}
\submitted{In Preparation for  ApJ}
\begin{document}
\pagenumbering{arabic}


\title{KELT-8\MakeLowercase{b}: A highly inflated transiting hot Jupiter and a new technique for extracting high-precision radial velocities from noisy spectra}
\author{
Benjamin J.\ Fulton\altaffilmark{1,20},
Karen A.\ Collins\altaffilmark{2,5},
B.\ Scott Gaudi\altaffilmark{4},
Keivan G.\ Stassun\altaffilmark{5,15},
Joshua Pepper\altaffilmark{3,5},
Thomas G.\ Beatty\altaffilmark{23},
Robert J. Siverd\altaffilmark{19},
Kaloyan Penev\altaffilmark{12},
Andrew W. Howard\altaffilmark{1},
Christoph Baranec\altaffilmark{1},
Giorgio Corfini\altaffilmark{8},
Jason D. Eastman\altaffilmark{10},
Joao Gregorio\altaffilmark{6},
Nicholas M. Law\altaffilmark{21},
Michael B. Lund\altaffilmark{3},
Thomas E. Oberst\altaffilmark{7},
Matthew T. Penny\altaffilmark{4,25},
Reed Riddle\altaffilmark{22},
Joseph E. Rodriguez\altaffilmark{3},
Daniel J.\ Stevens\altaffilmark{4},
Roberto Zambelli\altaffilmark{8},
Carl Ziegler\altaffilmark{21},
Allyson Bieryla\altaffilmark{10},
Giuseppe D'Ago\altaffilmark{24},
Darren L. DePoy\altaffilmark{18},
Eric L.\ N.\ Jensen\altaffilmark{13},
John F.\ Kielkopf\altaffilmark{2},
David W. Latham\altaffilmark{10},
Mark Manner\altaffilmark{14},
Jennifer Marshall\altaffilmark{18},
Kim K. McLeod\altaffilmark{16},
Phillip A.\ Reed\altaffilmark{17}
}
\altaffiltext{1}{Institute for Astronomy, University of Hawai`i, 2680 Woodlawn Drive, Honolulu, HI 96822, USA; email: bfulton@hawaii.edu} 
\altaffiltext{2}{Department of Physics and Astronomy, University of Louisville, Louisville, KY 40292, USA}
\altaffiltext{3}{Department of Physics, Lehigh University, Bethlehem, PA 18015, USA}
\altaffiltext{4}{Department of Astronomy, The Ohio State University, 140 West 18th Avenue, Columbus, OH 43210, USA}
\altaffiltext{5}{Department of Physics and Astronomy, Vanderbilt University, Nashville, TN 37235, USA}
\altaffiltext{6}{Atalaia Group and Crow-Observatory, Portalegre, Portugal}
\altaffiltext{7}{Department of Physics, Westminster College, New Wilmington, PA 16172, USA}
\altaffiltext{8}{Società  Astronomica Lunae, Castelnuovo Magra 19030, Via Montefrancio, 77 - Italy}
\altaffiltext{10}{Harvard-Smithsonian Center for Astrophysics, Cambridge, MA 02138 USA}
\altaffiltext{12}{Department of Astrophysical Sciences, Princeton University, Princeton, NJ 08544}
\altaffiltext{13}{Department of Physics and Astronomy, Swarthmore Col- legs, Swarthmore, PA 19081, USA}
\altaffiltext{14}{Spot Observatory, Nunnelly, TN 37137, USA}
\altaffiltext{15}{Department of Physics, Fisk University, Nashville, TN 37208, USA}
\altaffiltext{16}{Wellesley College, Wellesley, MA 02481, USA}
\altaffiltext{17}{Department of Physical Sciences, Kutztown University, Kutztown, PA 19530, USA}
\altaffiltext{18}{George P. and Cynthia Woods Mitchell Institute for Fundamental Physics and Astronomy, Texas A and M University, College Station, TX 77843-4242, USA}
\altaffiltext{19}{Las Cumbres Observatory Global Telescope Network, 6740 Cortona Drive, Suite 102, Santa Barbara, CA 93117, USA}
\altaffiltext{20}{NSF Graduate Research Fellow}
\altaffiltext{21}{Department of Physics and Astronomy, University of North Carolina at Chapel Hill, Chapel Hill, NC 27599, USA}
\altaffiltext{22}{Division of Physics, Mathematics, and Astronomy, California Institute of Technology, Pasadena, CA 91125, USA}
\altaffiltext{23}{Department of Astronomy and Astrophysics and Center for Exoplanets and Habitable Worlds, The Pennsylvania State Uni- versity, University Park, PA 16802}
\altaffiltext{24}{Istituto Internazionale per gli Alti Studi Scientifici (IIASS), Via G. Pellegrino 19, 84019, Vietri Sul Mare (SA), Italy}
\altaffiltext{25}{Sagan Fellow}

\begin{abstract}
We announce the discovery of a highly inflated transiting hot Jupiter discovered by the KELT-North survey. A global analysis including constraints from isochrones
indicates that the V = 10.8 host star (HD 343246) is a mildly evolved, G dwarf with $T_{\rm eff} = 5754_{-55}^{+54}$ K,
$\log{g} = 4.078_{-0.054}^{+0.049}$, $[Fe/H] = 0.272\pm0.038$, an inferred mass $M_{*}=1.211_{-0.066}^{+0.078}$ M$_{\odot}$, and radius
$R_{*}=1.67_{-0.12}^{+0.14}$ R$_{\odot}$. The planetary companion has mass $M_P = 0.867_{-0.061}^{+0.065}$ $M_{J}$, radius $R_P = 1.86_{-0.16}^{+0.18}$ $R_{J}$,
surface gravity $\log{g_{P}} = 2.793_{-0.075}^{+0.072}$, and density $\rho_P = 0.167_{-0.038}^{+0.047}$ g cm$^{-3}$. The planet is on a roughly circular orbit with semimajor
axis $a = 0.04571_{-0.00084}^{+0.00096}$ AU and eccentricity $e = 0.035_{-0.025}^{+0.050}$.
The best-fit linear ephemeris is $T_0 = 2456883.4803 \pm 0.0007$ BJD$_{\rm TDB}$ and $P = 3.24406 \pm 0.00016$ days.
This planet is one of the most inflated of all known transiting exoplanets, making it one of the few members of a class of extremely low density, highly-irradiated gas giants.
The low stellar $\log{g}$ and large implied radius are supported by stellar density constraints from follow-up light curves, plus an evolutionary and space motion analysis.
We also develop a new technique to extract high precision radial velocities from noisy spectra that reduces the observing time needed to confirm transiting planet candidates. This planet boasts deep transits of a bright star, a large inferred atmospheric scale height, and
a high equilibrium temperature of $T_{eq}=1675^{+61}_{-55}$ K, assuming zero albedo and perfect heat redistribution, making it one of the best targets for future atmospheric characterization studies.
\end{abstract}


{\it Facilities:} \facility{APF}, \facility{PO:1.5m (Robo-AO)}, \facility{Keck:I (HIRES)}

\section{Introduction}
\label{sec:intro}

Ground-based surveys for transiting exoplanets have been extraordinarily successful in detecting Jupiter-size planets orbiting very close to their host stars (hot Jupiters). These planets are some of the easiest to detect and characterize but also some of the most intrinsically rare, occurring around only 0.3 to 1.5\% of Sun-like stars \citep{Gould06,Cumming08,Bayliss11,Wright12,Howard12}. Both radial velocity (RV) surveys and transit surveys are heavily biased to detect these massive close-in objects yet we still know of only 162 examples\footnote{Planets with orbital periods shorter than 10 days and $\msini\geq0.5\mj$ based on a 2015 Mar 3 query of exoplanets.org} of hot Jupiters out of the thousands of currently known exoplanets.

Due to their short orbital periods and large transit and/or radial velocity (RV) signals, these planets make excellent laboratories to study planet formation and migration theories \citep{Hansen12, Ida08, Mordasini09}, atmospheric properties and composition \citep{Zhao14, Lockwood14, Sing15}, and even the rotational velocities of giant planets \citep{Snellen14}. Several mysteries remain unsolved for this population of exoplanets including the differences between the orbital distance distribution of close-in Jupiters orbiting metal rich and metal poor stars \citep{Dawson13}, the frequency of long-period companions in hot Jupiter systems \citep{Knutson14, Ngo15}, the source of large misalignments between the orbital axis and stellar spin axis \citep{Albrecht12, Dawson14}, and the reason that many hot Jupiters are inflated to extremely large radii \citep{Batygin11, Ginzburg15}. Each hot Jupiter discovery enhances our ability to explore these phenomena and study individual systems in exquisite detail.

RV surveys have discovered many non-transiting hot Jupiters around very bright stars, while the target stars of transit surveys tend to be much fainter. This is primarily due to the fact that only a narrow range of orbital parameters cause a planet to transit and many tens of thousands of stars must be observed in order to detect a single transiting planet. Since bright stars are distributed across a large area of the sky it is generally more efficient to observe small areas of the sky to a greater depth for transits of fainter stars. The Kilodegree Extremely Little Telescope survey \citep[KELT,][]{Pepper07} serves to bridge the gap in host star brightness between RV-detected and transit-detected planets by using extremely small aperture telescopes with large fields of view to observe nearly the entire sky. The survey is optimized to detect planets orbiting stars with V-band magnitudes between 8 and 10 but it has also been successful in detecting planets around slightly fainter stars as well \citep{Collins14}.

We report the discovery of a highly inflated ($R_p = \Rp$ \rj, $M_p = \Mp$ \mj) hot Jupiter orbiting the moderately bright V=10.84 G dwarf; HD 343246 (KELT-8 hereafter). KELT-8b has the 2nd largest radius and 7th lowest density among all transiting exoplanets, with only WASP-17b being larger \citep{Southworth12,Bento14}. The planet lies well above the \citet{Seager07} mass-radius relation for a pure hydrogen composition. In order to facilitate rapid and efficient RV confirmation of the planet we developed a new technique that saves significant telescope resources when collecting RV measurements using the iodine technique \citep{Butler96b}. In \S \ref{sec:observations} we describe our discovery and follow-up observations. We describe our spectroscopic analysis of the host star and summarize the inferred stellar properties in \S \ref{sec:stellar_props}. A close companion stellar object that was ultimately found to be a background contaminant is described in \S \ref{sec:companion}. We outline our new technique for RV extraction in \S \ref{sec:synthRV}. A global analysis of the photometric and RV data is presented in \S \ref{sec:characterization}, and we conclude with a discussion in \S \ref{sec:discussion} and summary in \S \ref{sec:summary}.

\section{Observations}
\label{sec:observations}

\subsection{KELT Photometry}

The KELT-North survey telescope consists of an Apogee AP16E imager 
(4K x 4K CCD with 9$\mu$m pixels) and an 80mm Mamiya 645 camera lens
(42 mm diameter, f/1.9) behind a Kodak Wratten \#8 red-pass filter. This
setup achieves a $26^\circ \times 26^\circ$ field of view with roughly
23'' pixel$^{-1}$. For a more complete description of hardware and operations,
see \citet{Pepper07}.

Raw KELT science images are dark-subtracted and flat-fielded using
standard methods and then reduced using a heavily-modified implementation
of the ISIS image subtraction package \citep{Alard98,Alard00}
coupled with our own high-performance background-removal routines. Stars
are identified for extraction using the standalone DAOPHOT II PSF-fitting
software package \citep{Stetson87, Stetson90}. To reduce systematic errors, extracted
light curves are processed with the trend filtering algorithm \citep[TFA;][]{Kovacs05}
prior to period search. A more complete description
of the data reduction pipeline and candidate selection process are available
in \citet{Siverd12}.

\begin{figure*}[ht] 
   \centering
\includegraphics[width=0.75\textwidth]{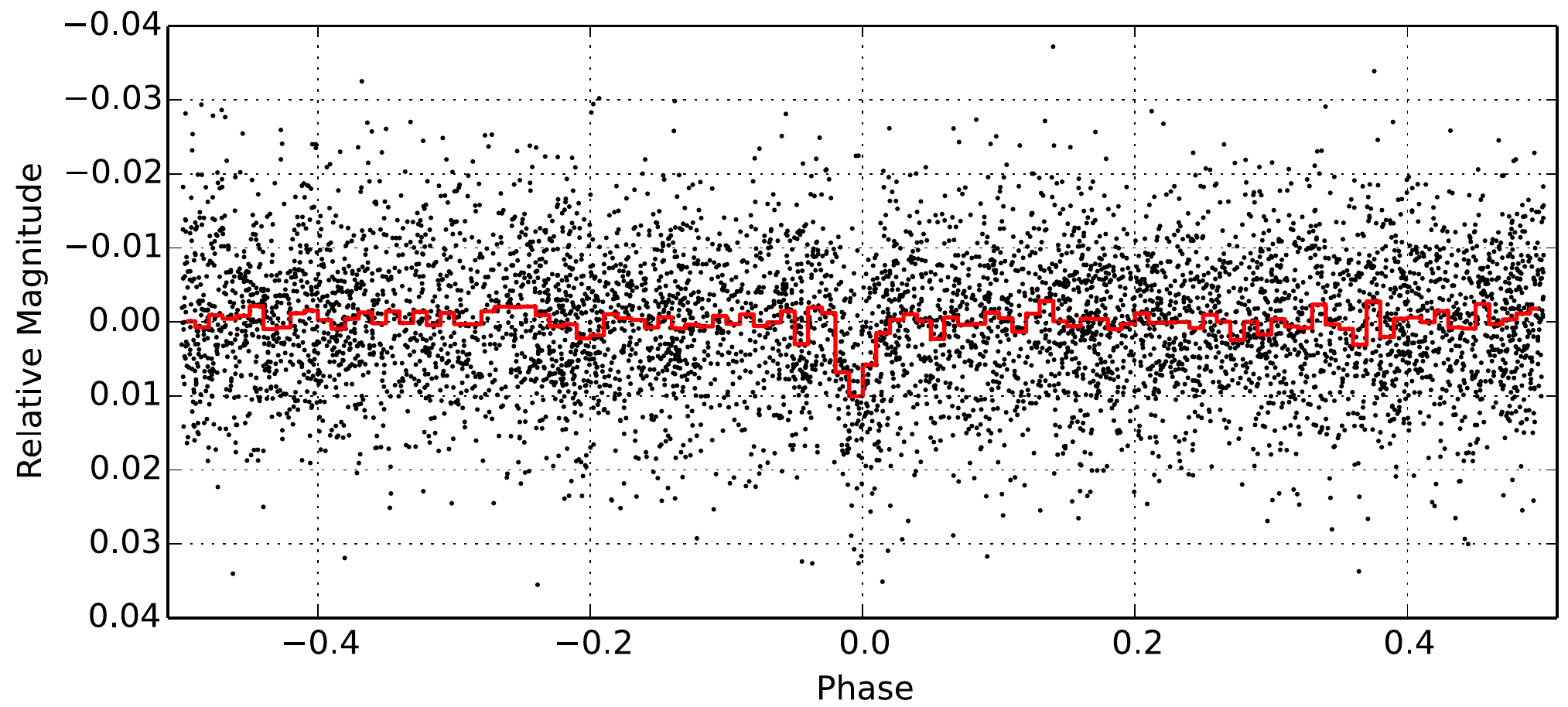}
   \caption{Discovery light curve of KELT-8b from the KELT-North telescope. The photometry is phase-folded to the ephemeris listed in Table \ref{tab:params}. The red line is the same data binned in intervals of width 0.01 in orbital phase ($\approx$46.7 minutes).}
   \label{fig:keltphot}
\end{figure*}

KELT-8 resides in KELT-North field 11, centered on 
($\alpha = 19^h 26^m 48^s, \; \delta = +31^\circ 39' 56''$; J2000).
The data set consists of 5978 images acquired between 30 May, 2007 and
14 June, 2013. The full discovery light curve, phased to the KELT-8b transit
ephemeris, is shown in Figure \ref{fig:keltphot}.

\subsection{Follow-up Photometry}

\begin{figure}[ht] 
   \centering
\includegraphics[width=0.46\textwidth]{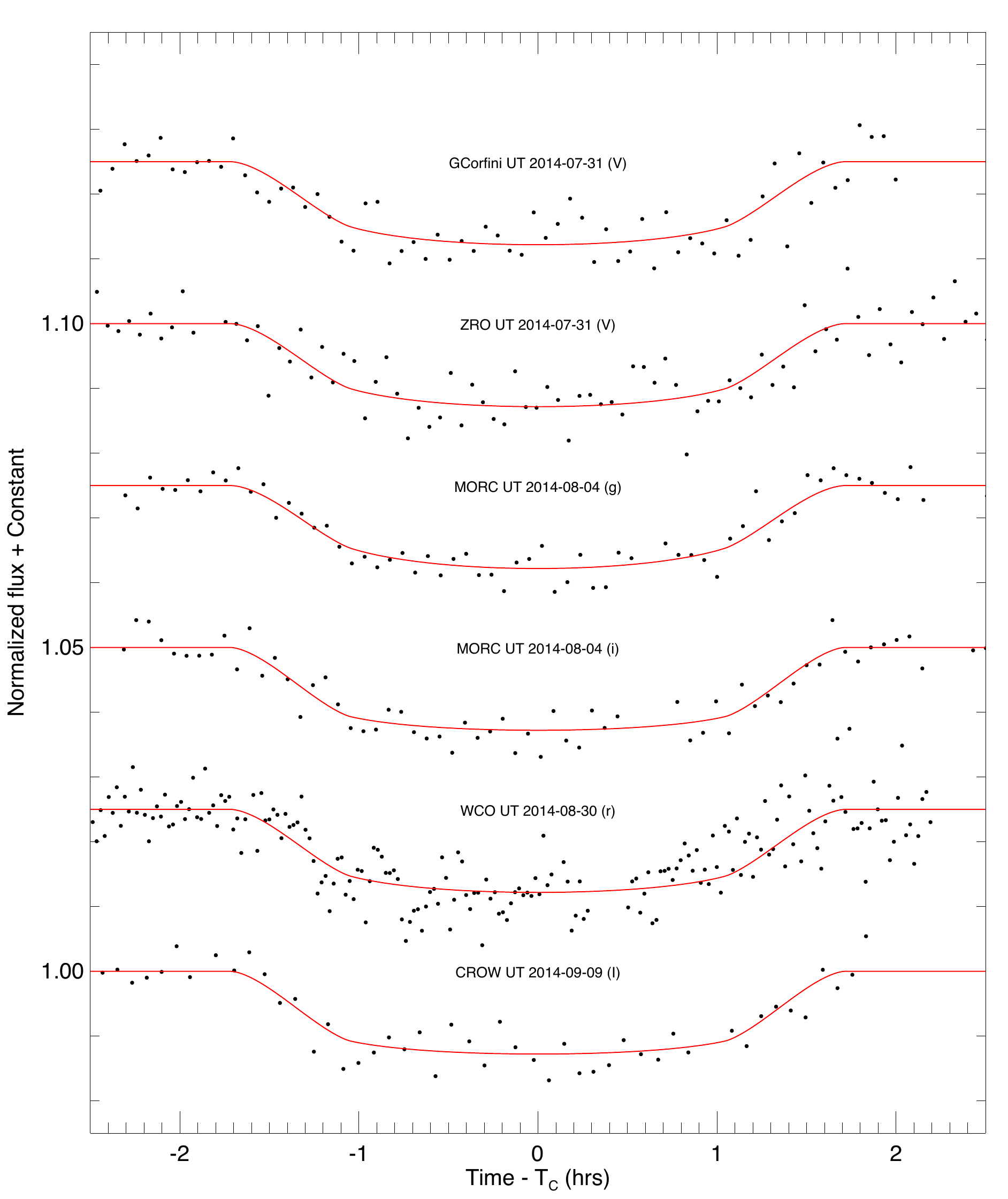}
\includegraphics[width=0.46\textwidth]{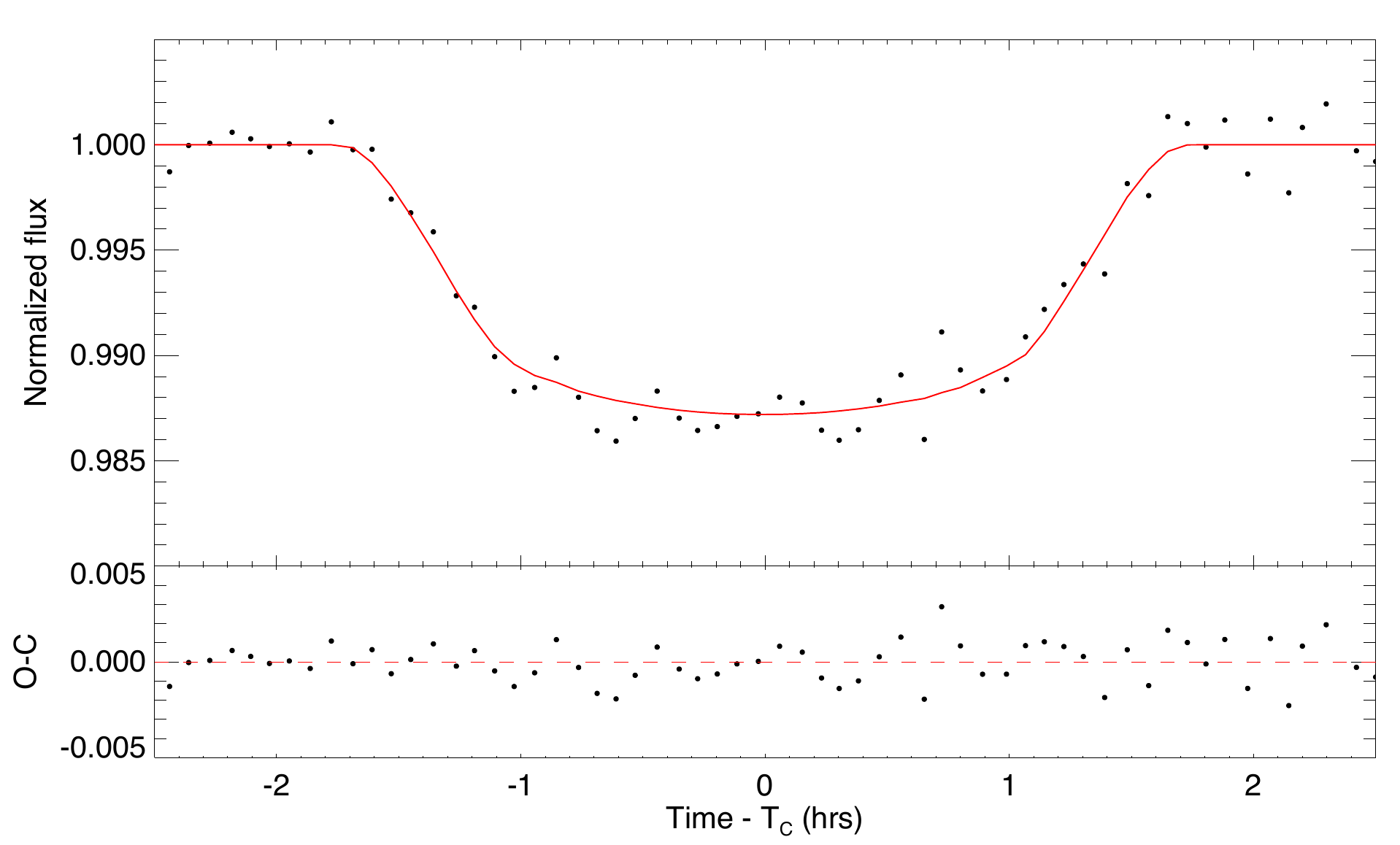}
   \caption{\emph{Top:} Follow-up photometry of KELT-8b primary transits. The source, date, and filter for each transit is annotated. The best-fit models are shown in red. \emph{Bottom:} All follow-up transit light curves combined and binned in 5 minute intervals. This light curve is not used in the analysis and is shown simply to demonstrate the combined photometric precision of the follow-up light-curves.}
\label{fig:transits}
\end{figure}

We acquired follow-up time-series photometry of KELT-8 to better determine the transit shape and to check for a consistent transit depth across the optical filter bands. To schedule follow-up photometry, we used the {\tt Tapir} software package \citep{Jensen13}. We obtained full primary transits in multiple bands between July 2014 and September 2014. Figure \ref{fig:transits} shows all the primary transit follow-up light curves assembled. A summary of the follow-up photometric observations is shown in Table \ref{tab:photobs}. We find consistent $R_{\rm P}/R_\star$ ratios in the $g'$, $r'$, $i'$, $V$, and $I_C$ filters
after compensating for blending with a nearby neighbor in some observations (see below), helping to rule out false positives due to blended eclipsing binaries. The lower panel of Figure \ref{fig:transits} shows all primary transit follow-up light curves from Figure \ref{fig:transits} combined and binned in 5 minute intervals. This combined and binned light curve is not used for analysis, but rather to show the overall statistical power of the follow-up photometry.

AstroImageJ\footnote{http://www.astro.louisville.edu/software/astroimagej/} (AIJ) (K.~A.~Collins \& J.~F.~Kielkopf 2015, in preparation) was used to calculate the differential photometric data from all calibrated image sequences. AIJ is a general purpose image processing package, but is optimized for processing time-series astronomical image sequences.

\subsubsection{MORC}
We observed one complete KELT-8b transit on UT 2014-08-04 from the 0.6 m Moore Observatory RCOS (MORC) telescope, operated by the University of Louisville. MORC is equipped with an Apogee U16M 4K $\times$ 4K CCD giving a 26' $\times$ 26' field of view and 0.39 arcseconds pixel$^{-1}$. The transit was observed pseudo-simultaneously in two filters by alternating between the $g'$ and $i'$ filters from one exposure to the next. The filter change is made during CCD read-out to minimize loss of time on the the sky.
Since KELT-8b has a nearby neighbor, telescope defocusing was minimal to provide better separation of the flux from the two stars on the detector.

\subsubsection{GCorfini}
We observed one full transit on UT 2014-07-31 from Giorgio Corfini's home observatory (GCorfini) in Lucca, Italy. The observations were obtained in V-band using a 0.2 m Newtonian telescope and a SBIG STT-6303 ME CCD 1536 $\times$ 1024 pixel camera, giving a 59' $\times$ 39' field of view.
No defocusing was applied, but the target and neighbor star were severely blended on the detector due to seeing. A photometric aperture was chosen that included both KELT-8 and its neighbor. The resulting light curve was corrected for blending as described in \S \ref{sec:characterization}.

\subsubsection{ZRO}
We observed one full transit in the $V_C$ filter on UT 2014-07-31 at Roberto Zambelli's personal observatory (ZRO). The telescope was not defocused and both stars were cleanly resolved from each other. The observations were obtained using a Meade LX 200 ACF GPS 12" telescope, a f/6.3 focal reducer, and an SBIG ST8 XME 1530 $\times$ 1020 CCD with a pixel scale of 0.92 arc seconds pixel$^{-1}$ and a 23' $\times$ 15' field of view.

\subsubsection{WCO}
We observed a full transit in the $r'$ band at Westminster College Observatory (WCO) in Pennsylvania on UT 2014-08-30. The observations were obtained using a Celestron 0.35 m C14 telescope with an SBIG STL-6303E 3072 $\times$ 2048 CCD, giving a 24' $\times$ 16' field of view and 0.5 arcseconds pixel$^{-1}$. The data were collected with the telescope in focus and the two stars were resolved from each other.

\subsubsection{CROW}
We observed one full transit in the $I_C$ band at Canela's Robotic Observatory (CROW) in Portugal on UT 2014-09-09. The observations were obtained using a 0.3 m LX200 telescope with an SBIG ST-8XME 1530 $\times$ 1020 CCD, giving a 28' $\times$ 19' field of view and 1.11 arcseconds pixel$^{-1}$. No telescope defocus was applied, but KELT-8 and the neighbor were blended on the detector due to seeing. A photometric aperture was selected to encircle the flux from both stars. The resulting light curve was corrected for blending as described in \S \ref{sec:characterization}.

\begin{deluxetable}{lcrcrccc}[ht]
\tabletypesize{\footnotesize}
\tablecaption{Summary of Photometric Observations}
\tablehead{ 
    \colhead{Telescope}               & \colhead{UT}  & \colhead{\#} &  \colhead{Filter} & \colhead{Cycle\tablenotemark{a}} & \colhead{RMS\tablenotemark{b}} & \colhead{PNR\tablenotemark{c}} & \colhead{$\alpha$\tablenotemark{d}} \\
    \colhead{}  & \colhead{(2014)}            & \colhead{Obs}    & \colhead{Band}    & \colhead{(sec)} & \colhead{($10^{-3}$)} & \colhead{($\frac{10^{-3}}{\rm min}$)} & \colhead{} 
}
\startdata

GCorfini & 7-31 & 73   & $V$   & 243 & 3.0 & 6.0 & 1.1 \\
ZRO      & 7-31 & 110 & $V$   & 216 & 3.5 & 6.6 & 1.9\\
MORC   & 8-04  & 65  & $g'$   & 127 & 2.1 & 3.0 & 2.2\\
MORC   & 8-04  & 62  & $i'$    & 127 & 2.3 & 3.3 & 2.6\\
WCO    & 8-30  & 209 & $r'$    & 81 & 3.8 & 4.4 & 3.1\\
CROW  & 9-09  & 49  & $I_C$ & 332 & 2.3 & 5.4 & 1.3\\

\enddata
\tablenotetext{a}{Cycle time in seconds, calculated as the mean of exposure time plus dead time during periods of back-to-back exposures. The MORC $g'$ and $i'$ exposures were alternating, so cycle time is calculated for exposures in both filters combined.}
\tablenotetext{b}{RMS of residuals from the best fit model in units of $10^{-3}$ .}
\tablenotetext{c}{Photometric noise rate in units of $10^{-3}$ minute$^{-1}$, calculated as RMS/$\sqrt{\Gamma}$, where RMS is the scatter in the light curve residuals and $\Gamma$ is the mean number of cycles (exposure time and dead time) per minute during periods of back-to-back exposures (adapted from \citealt{Fulton11}).}
\tablenotetext{d}{Scaling factor applied to the uncertainties to ensure that the best fitting model has \chisq=1 (see \S \ref{sec:characterization}).}

\label{tab:photobs}
\end{deluxetable}
\vspace{10pt}

\subsection{High-resolution Spectroscopy}
\label{sec:RV}

We collected a total of 13 RV measurements of KELT-8 using the Levy high-resolution optical spectrograph mounted at the Nasmyth foci of the Automated Planet Finder Telescope (APF)  at Lick observatory \citep{Vogt14a, Radovan14, Burt14}. The measurements were collected between UT 2014 August 15 and UT 2014 November 9 and are presented in Table \ref{tab:rv}. We observed this star using the $2\times8\arcsec$ slit for a spectral resolution of R $\approx$ 80,000. We pass the starlight through a cell of gaseous iodine which serves as a simultaneous wavelength and point spread function (PSF) reference \citep{Marcy92}. Relative radial velocities are calculated by tracing the doppler shift of the stellar spectrum with respect to the dense forest of iodine lines using a forward modeling technique described in \citet{Butler96b}. Traditionally, a high signal-to-noise iodine-free observation of the same star is deconvolved with the instrumental PSF and used as the stellar template in the forward modeling process. However, in this case the star is too faint to collect the signal-to-noise needed for reliable deconvolution in a reasonable amount of time on the APF. Instead we simulate this observation by using the SpecMatch software (Petigura et al. 2015 in prep.) to construct a synthetic template from the \citet{Coelho14} models and best fit stellar parameters. This process is described in more detail in Section \ref{sec:synthRV}.

\begin{deluxetable}{ccccc}
\tabletypesize{\footnotesize}
\tablecaption{Radial Velocities of KELT-8}
\tablewidth{245pt}
\tablehead{ 
    \colhead{\bjdtdb}               & \colhead{RV}  & \colhead{$\sigma_{\rm RV}$} &  \colhead{BS} & \colhead{$\sigma_{\rm BS}$} \\
    \colhead{(-- 2440000)}  & \colhead{(\mse)}            & \colhead{(\mse)}       & \colhead{(\mse)} & \colhead{(\mse)}
}
\startdata

  16884.7240398  &  -52.700  &  13.698 & 18.194  &  14.832 \\
  16886.7223760  &  16.089   &   11.487 & 60.348  &  22.517 \\
  16888.6894588  &   54.770  &  11.598 &  21.460  &  21.508 \\
  16890.6920118  &  -85.923  &  11.536 &  104.118 &  56.715 \\
  16891.6849308  &   39.139  &  11.894 &  -18.528  &  26.524 \\
  16892.6862628  &  120.081  &  10.732 &  15.611   &  12.552 \\
  16894.6776168  &  -30.453  &  11.729  &  -18.657  &  35.822 \\
  16895.6940738  &   98.475  &  11.258   &  18.095  &  18.990 \\
  16900.7203146  &  -99.117  &  11.290  &  -3.809  &  25.208 \\
  16903.7467735  &  -104.461  &  12.930 &  67.604  &  41.392 \\
  16905.6816215  &   91.985  &  13.023  &  6.130  &  15.736 \\
  16906.7226164  &  -90.377  &  12.995  & 117.227 &  77.109 \\
  16970.6342444  &   55.520  &  20.228  &  73.214  &  34.466 \\

\label{tab:rv}
\end{deluxetable}
\vspace{10pt}

\subsection{Adaptive Optics Imaging}
\label{sec:AO}
We acquired visible-light adaptive optics images of KELT-8 using the Robo-AO system \citep{Baranec2013, Baranec2014} on the 60-inch Telescope at Palomar Observatory. On UT 2015 March 8, we observed KELT-8 in the Sloan-$i'$ filter as a sequence of full-frame-transfer detector readouts at the maximum rate of 8.6 Hz for a total of 120 s of integration time. The individual images are corrected for detector bias and flat-fielding effects before being combined using post-facto shift-and-add processing using KELT-8 as the tip-tilt star with 100\% frame selection to synthesize a long-exposure image \citep{Law2014}.

\begin{figure}[htbp] 
   \centering
\includegraphics[width=3in]{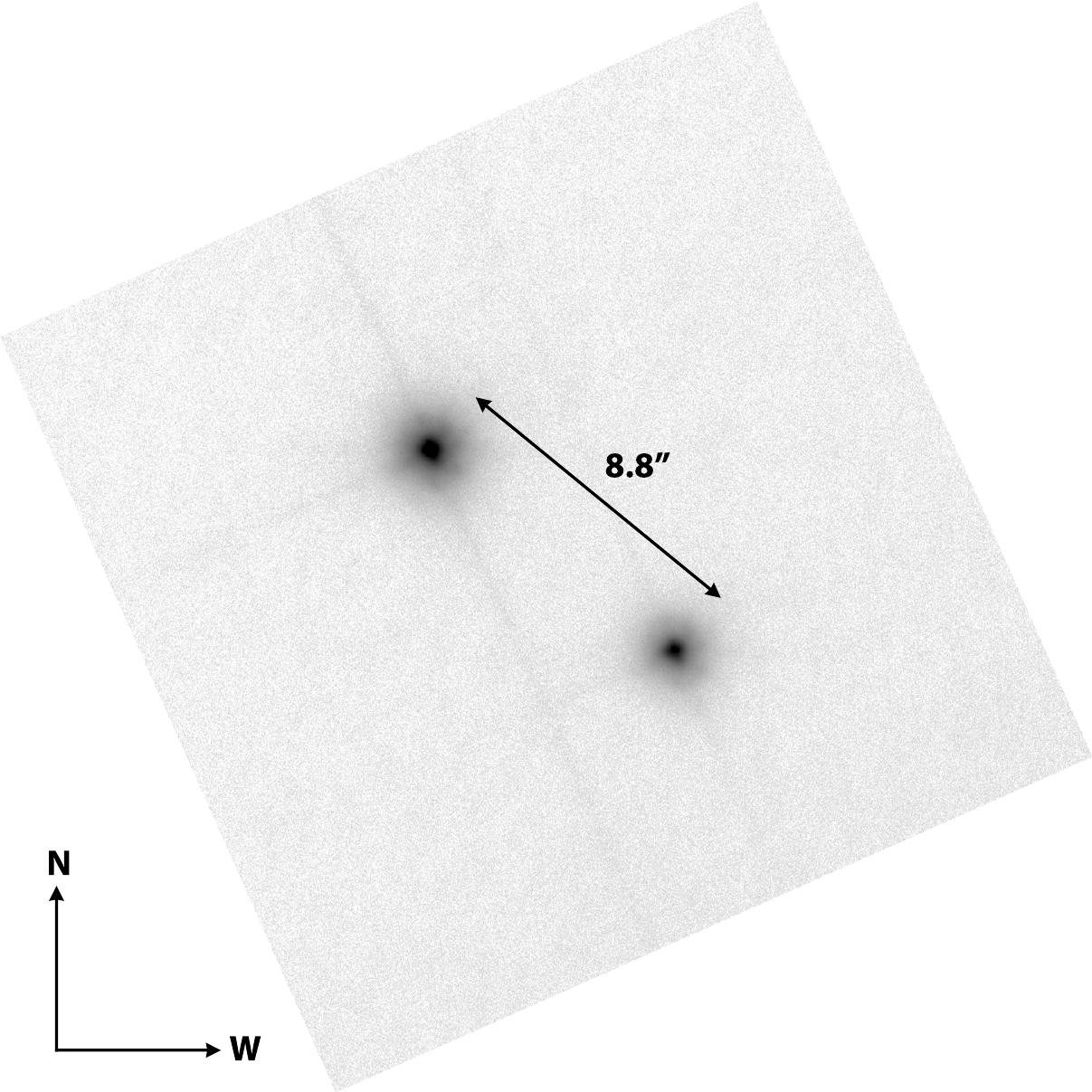}
\includegraphics[width=3in]{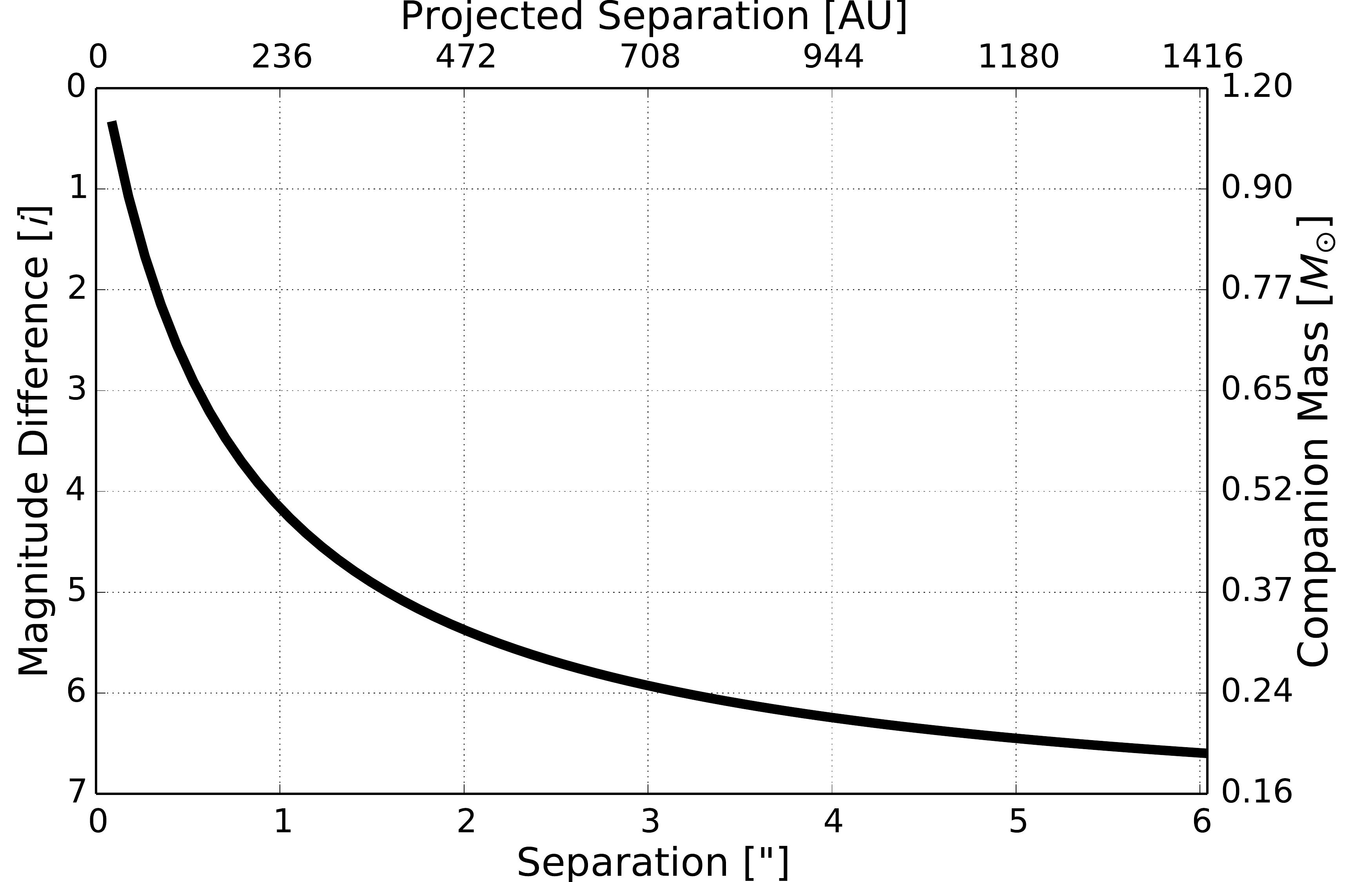}
   \caption{\emph{Top:} Robo-AO image of KELT-8 showing the background giant star 8\farcs8 to the SW of KELT-8. \emph{Bottom:} Five $\sigma$ contrast limit near KELT-8 derived from the Robo-AO image.}
   \label{fig:ao}
\end{figure}

We calculate the 5 $\sigma$ contrast limit as a function of angular separation by first determining the background noise level for concentric rings of width equal to the FWHM moving outward from the primary star.  A simulated companion (a dimmed cutout of the primary star) is inserted into the closest ring, with a random PA.  An auto-companion detection code then searches for the simulated companion.  If it is found, the companion is dimmed further and reinserted. This is continued until the companion is not found with a confidence greater than 5 $\sigma$. We repeat this process for each annulus, and fit the sparse measurements with a function of the form $a - b/(r - c)$, where $a$, $b$, and $c$ are free parameters in the fit and r is the radius from the target star. We convert contrast limits to mass limits using the models of \citet{Baraffe02}. The resulting contrast and mass limits are presented in Figure \ref{fig:ao}

\section{Stellar Properties}
\label{sec:stellar_props}
\subsection{Keck/HIRES Spectroscopy}
In order to obtain precise values for the stellar parameters we collected a moderate signal-to-noise iodine-free observation using the HIRES spectrograph on the Keck I telescope \citep{Vogt94}. We measured the effective temperature (\teff), surface gravity (\logg), iron abundance (\feh), and rotational velocity of the star (\vsini) using the tools available in the SpecMatch software package (Petigura et al. 2015 in prep). We first correct the observed wavelengths to be in the observer's rest frame by cross-correlating a solar model with the observed spectrum. Then we fit for \teff, \logg, \feh, \vsini, and the instrumental PSF using the underlying differential-evolution Markov Chain Monte Carlo \citep[DE-MCMC,][]{Braak06} machinery of ExoPy \citep{Fulton13}. At each point in the MCMC chains a synthetic spectrum is created by interpolating the \citet{Coelho14} grid of stellar models for a set of \teff, \logg, \feh\ values and solar alpha abundance. We convolve this synthetic spectrum with a rotational plus mactroturbulance broadening kernel using the prescriptions of \citet{Valenti05} and \citet{Hirano11}. Finally, we perform another convolution with a gaussian kernel to account for the instrumental PSF and the synthetic spectrum is compared with the observed spectrum using $\chi^2$ to assess the goodness of fit. The priors are uniform in \teff, \logg, and \feh, but we assign a gaussian prior to the instrumental PSF that encompasses the typical variability in the PSF width caused by seeing changes and guiding errors. Five echelle orders of the spectrum are fit separately and the resulting posterior distributions are combined before taking the median values for each parameter.

Parameter uncertainties are estimated as the scatter in spectroscopic parameters given by SpecMatch relative to the values in \citet{Valenti05} for 352 of the stars in their sample and 76 stars in the \citet{Huber12} asteroseismic sample. Systematic trends in SpecMatch values as a function of \teff, \logg, and \feh\ relative to these benchmark samples are fit for and removed in the final quoted parameter values. More details about the benchmark comparison and calibrations will be available in Petigura et al. (2015, in prep). These spectroscopic parameters are then used as input priors for the global fit discussed in \S \ref{sec:characterization}. The best fit spectroscopically-determined stellar parameters are $\logg=4.23\pm0.08$, $\feh=0.272\pm0.038$, and $\teff=5754^{+54}_{-55}$ K.

The $\logg$ derived from spectroscopy is only marginally consistent with the $\logg$ derived from the global fit ($4.078_{-0.054}^{+0.049}$). The $\logg$ value measured from the transit light curves is likely more reliable because $\logg$ can be a difficult quantity to measure from high-resolution spectroscopy alone. We perform a global analysis both with and without the spectroscopic $\logg$ as a prior (see \S \ref{sec:characterization}).

We also collected a single exposure with the iodine cell in the light path with a slightly higher S/N ratio to establish a long term baseline for possible followup of this target with Keck in the future. We used this observation to extract the \lrphk\ stellar activity metric. We find that \lrphk =-5.108 which indicates that the star very chromospherically quiet.


\subsection{Spectral Energy Distribution}
We estimate the distance and reddening to KELT-8 by fitting the B-band through WISE W4-band spectral energy distribution (SED) to the \citet{Kurucz1979} stellar atmosphere models. We also found a GALEX NUV measurement in the literature \citep{Bianchi11}, however this measurement shows a large excess in flux relative to the SED compatible with all other measurements. Since the PSF of GALEX in the NUV band is 4.9" we suspect the GALEX measurement is contaminated by the visual companion 8\farcs8 away and we omit this measurement from the fit. The WISE measurements also have a large PSF relative to the separation between the two components, but we do not find any evidence for significant contamination. We fix \teff, \logg, and \feh\ to the best fit values in Table \ref{tab:params} and leave distance ($d$) and reddening ($A_V$) as free parameters. We find best-fit parameters of $A_V=0.15\pm0.06$ magnitudes and $d=236\pm9$ pc with a reduced $\chi^2=3.7$ (considering only the flux uncertainty and not the uncertainty in the model). This agrees quite well with the distance of 233 pc estimated by \citet{Pickles10}.

\begin{figure}[h] 
   \centering
\includegraphics[width=3in]{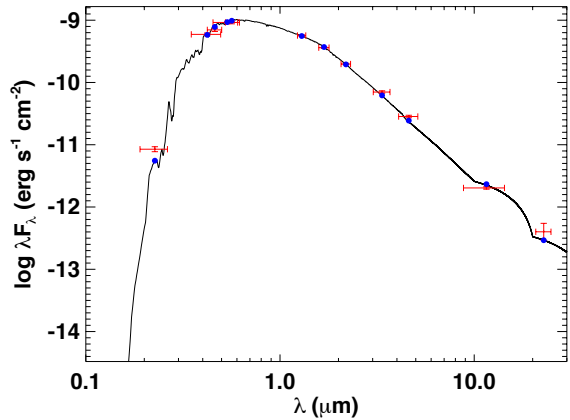}
   \caption{SED fit to KELT-8. The red crosses are the photometry measurements. The vertical errors are the 1 $\sigma$ photometric uncertainties, and the horizontal error bars are the effective widths of the passbands. The leftmost point (GALEX NUV) was omitted from the fit due to possible contamination from the neighboring star at 8\farcs8 separation. The black line is the best-fit stellar atmosphere model and the blue points are the predicted passband-integrated fluxes of the model corresponding to our observed photometric bands. }
   \label{fig:sed}
\end{figure}

\subsection{Evolutionary State}
\begin{figure}[h] 
   \centering
\includegraphics[width=3in]{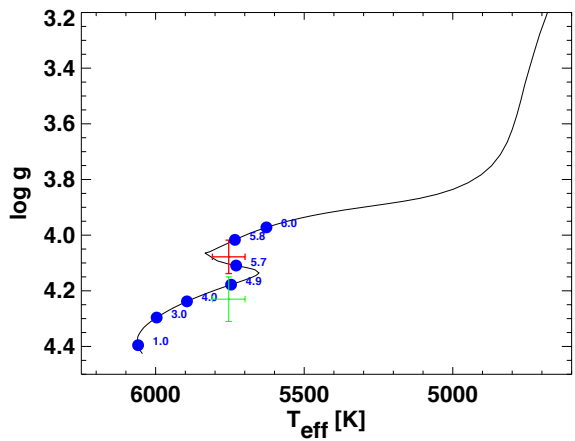}
   \caption{Yonsei-Yale isochrone fit for KELT-8. The solid black line is the isochrone for a star with the mass and metallically of KELT-8 listed in Table \ref{tab:params}. The red cross is the position of KELT-8, and the blue points label various ages along the evolutionary track. The green cross is the position of KELT-8 using the spectroscopically-determined \logg. The ``blue hook" is the kink in the track where the star's \teff\ increases while \logg\ decreases for a short amount of time between an age of 4.9 and 5.8 Gyr. }
   \label{fig:evo}
\end{figure}

We estimate the age of KELT-8 by fitting Yonsei-Yale isochrones to the values of \teff, \logg, and \feh\ given in Table \ref{tab:params}. We fix the stellar mass to the value of $M_{*}=1.211$ \msun listed in Table \ref{tab:params}. Our best fit stellar parameters indicate that the star happens to fall on a very rapid part of the evolutionary track, the so-called ``blue hook" just prior to crossing the Hertzsprung gap (see Figure \ref{fig:evo}).
In principle, this would allow for a very precise measurement of the stellar age, however, the measurement errors on \teff\ and \logg\ don't allow us to locate the star's exact position within the blue hook. We conservatively quote an age range of 4.9-5.8 Gyr that spans the entire blue hook.
We note that the uncertainty on the age likely does not follow a Gaussian distribution because the rapid evolution through the blue hook means that stars on the blue hook should be rare.
It is much more likely that it falls on one side or the other of the blue hook simply because stars spend such a small fraction of the lifetimes on the blue hook. A more detailed evolutionary analysis that includes rotation, chemical composition, and a full consideration of all relevant priors is needed to determine the exact evolutionary state of KELT-8.

\begin{deluxetable*}{lccc}
\tabletypesize{\footnotesize}
\tablecaption{Adopted Stellar Properties of KELT-8}
\tablewidth{0pt}
\tablehead{
  \colhead{Parameter (units)}   & 
  \colhead{Value} &
  \colhead{Source} &
  \colhead{Reference}
}
\startdata
Names &  HD 343246 & SIMBAD & \\
            &  GSC 02109-00049 & SIMBAD & \\
            &  2MASS J18531332+2407385 & SIMBAD & \\
            &  TYC 2109-49-1 & SIMBAD & \\
            &  WDS 18533+2407 & WDS & \citet{Hartkopf13} \\
Spectral type ~~~~~~& G2V  & this work & \\
RA (J2000) & 18 53 13.3216   & Tycho-2  & \citet{Hog00} \\  
DEC (J2000) & +24 07 38.603 & Tycho-2 & \citet{Hog00} \\  
U\tablenotemark{a} (\kms) & $20.6 \pm 1.7$ & this work & \\
V (\kms) & $-32.6 \pm 1.2$ & this work & \\
W (\kms) & $0.7 \pm 1.5$ & this work & \\
RA proper motion (mas/yr)  & $-13.7\pm1.4$  & Tycho-2 & \citet{Hog00}  \\
DEC proper motion (mas/yr)  & $-29.5\pm1.3$  & Tycho-2 & \citet{Hog00}  \\
Systemic velocity (\kms) & -31.2 $\pm$ 0.3 & this work & \\

NUV (mag)  &  16.884 $\pm$ 0.1\tablenotemark{b}  & GALEX & \citet{Bianchi11}  \\

$B_{T}$ (mag) & 11.713 $\pm$ 0.057  & Tycho-2 & \citet{Hog00}  \\
$V_{T}$ (mag)  & 10.925 $\pm$ 0.048 &  Tycho-2 & \citet{Hog00}  \\

$B$ (mag) & 11.545 $\pm$ 0.065  & ASCC & \citet{Kharchenko01}  \\
$V$ (mag)  & 10.833 $\pm$ 0.054 &  ASCC & \citet{Kharchenko01}  \\

$J$ (mag)  & 9.586 $\pm$ 0.026 & 2MASS  & \citet{Cutri03}  \\
$H$ (mag)  & 9.269 $\pm$ 0.032 & 2MASS  & \citet{Cutri03} \\
$K$ (mag)   & 9.177 $\pm$ 0.021 & 2MASS & \citet{Cutri03}  \\

$W1$ (mag)  & 11.664 $\pm$ 0.05 & WISE  & \citet{Cutri14}  \\
$W2$ (mag)  & 12.302 $\pm$ 0.05 & WISE  & \citet{Cutri14} \\
$W3$ (mag)   & 14.17 $\pm$ 0.05 & WISE & \citet{Cutri14}  \\
$W4$ (mag)   & 15.193 $\pm$ 0.339 & WISE & \citet{Cutri14}  \\


\vsini (km\,s$^{-1}$) &  $3.7\pm1.5$  & this work & \\
Age (Gyr)  &  $5.4^{+0.4}_{-0.5}$  &  this work  & \\

Distance (pc)  &  $236\pm9$  &  this work  &  \\
$A_{V}$  &  $0.15\pm0.06$  &  this work  & \\

\lrphk & -5.108 & this work & \\
\enddata

\tablenotetext{a}{Positive U is in the direction of the Galactic Center.}
\tablenotetext{b}{Likely contaminated by the neighbor star 8\farcs8 away}
\label{tab:stellar_params}
\end{deluxetable*}
\vspace{20pt}

\subsection{Unrelated visual companion}
\label{sec:companion}

\begin{figure}[htbp] 
   \centering
\includegraphics[width=3.5in]{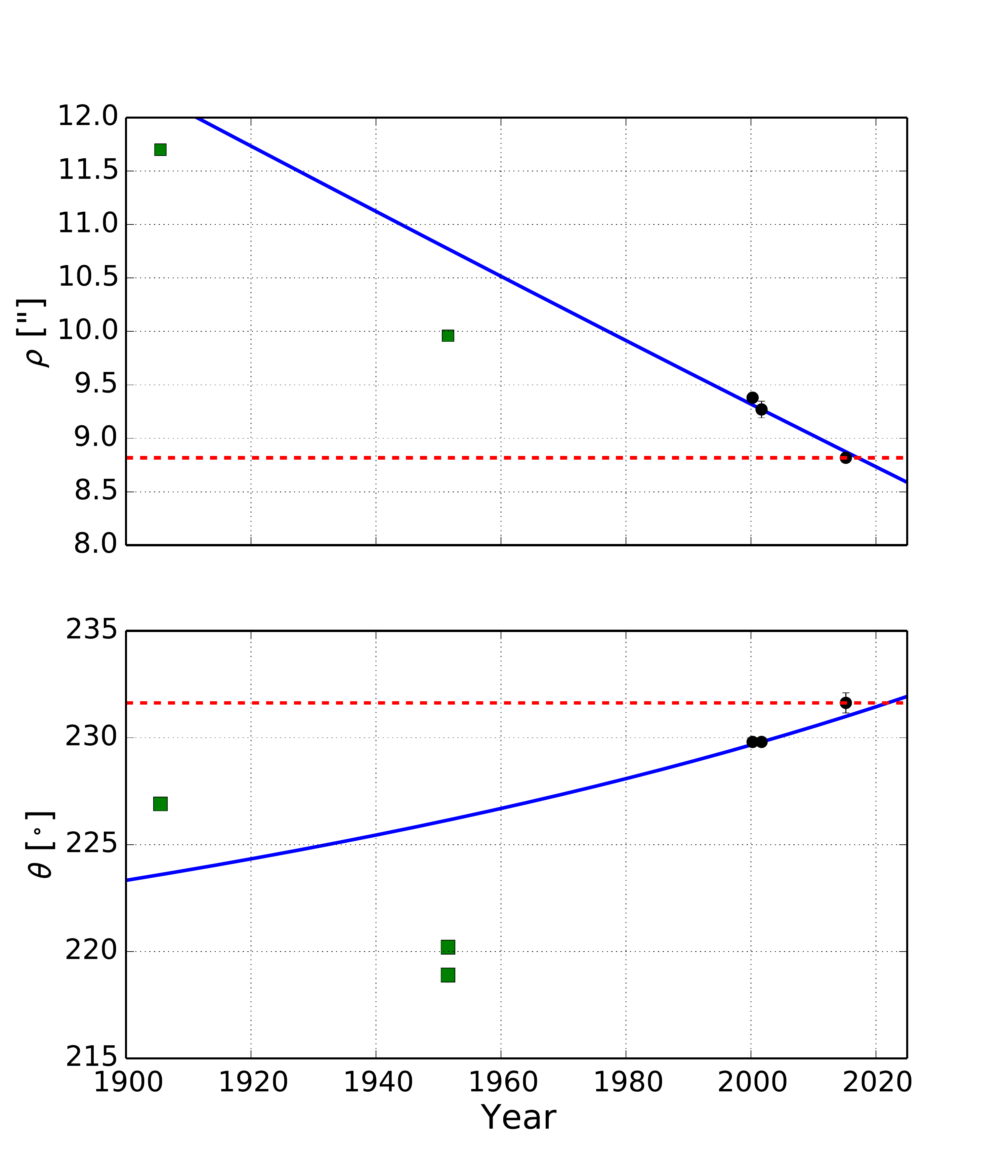}
   \caption{Astrometry measurements of KELT-8. The green squares were measured using astrometric eyepieces or photographic plates. The black points were measured using CCD data and the most recent point is from our Robo-AO observation. The blue lines are the predicted change in separation (upper panel) and position angle (lower panel) assuming the star 8\farcs8 to the SW of KELT-8 is in the background with negligible proper motion and KELT-8 has a constant proper motion with the values given in Table \ref{tab:stellar_params}. The dashed black line is a model assuming that the two components are physically bound and the orbital motion of the pair is negligible during the span of observations.}
   \label{fig:motion}
\end{figure}

KELT-8 was identified as a visual binary in 1905 by French astronomer Abel Pourteau \citep{Pourteau1933}. Figure \ref{fig:ao} shows KELT-8 and the fainter visual companion 8\farcs8 to the SW clearly resolved. In addition to the discovery astrometry, further astrometric measurements were made in 1951 \citep{Gellera1984, Gellera1990}, 2000 \citep{Cutri03}, and 2001 \citep{Hartkopf03}.  These measurements all show a trend of decreasing separation and increasing position angle, although the scatter in the pre-CCD measurements is large (see Figure \ref{fig:motion}). We also measured astrometry from our Robo-AO data and find that it is consistent with this trend, especially if only the CCD measurements are considered. This is consistent with KELT-8 moving with the constant proper motions given in Table \ref{tab:stellar_params} and the companion being stationary in the background.

We also collected an APF spectrum of the secondary component and used the SpecMatch pipeline to extract stellar parameters. We cross-correlated the best-fitting model with the observed spectrum after correcting for the Earth's barycentric motion to determine an absolute systemic velocity. SpecMatch returned $\teff=4702\pm60$, $\logg=2.77\pm0.08$, and $\feh=0.01\pm0.04$ which is consistent with the companion being a G or K giant with an intrinsic luminosity much greater than that of KELT-8. Since this component appears approximately one magnitude fainter than KELT-8 in V band it must be in the background. In addition, the systemic RV of the secondary of $34\pm1$ \kms is highly discrepant with the velocity of KELT-8 (-31.2 $\pm$ 0.3 \kse). These lines of evidence all suggest that the two stars are physically unrelated.

\subsection{UVW Space motion}
\label{sec:UVW}

We evaluate the motion of KELT-8 through the galaxy. The absolute radial velocity measured from the APF spectra of KELT-8 is $-31.2\pm0.3$ km s$^{-1}$, and the proper motions from Tycho-2 are $-13.7\pm1.4$ and $-29.5\pm1.3$ milli-arcseconds per year in RA and DEC respectively. These values transform to U,V,W space motions of $20.6 \pm 1.7$, $-32.6 \pm 1.2$, $0.7 \pm 1.5$ km s$^{-1}$ respectively, making KELT-8 an unambiguous member of the thin disk stellar population \citep{Bensby03}. Since it is not a halo star this further supports our adoption of the stellar $\logg$ derived from the transit fit which gives stellar parameters (particularly metallicity) that are much more common for thin disk stars \citep{Ivezic12}.

\section{Synthetic Template Radial Velocities}
\label{sec:synthRV}

Iodine-based radial velocities are traditionally derived using a forward modeling process that requires three primary ingredients; a template spectrum of the star with the iodine out of the light path, an ultra-high resolution spectrum of the iodine absorption for the particular cell in use, and a description of the instrumental PSF \citep{Butler96b}. The stellar template and the iodine spectrum are multiplied together then convolved with the PSF to match the observed spectrum taken through iodine. The stellar template is usually an observation of the same star taken at 2-3 times higher signal-to-noise (S/N) ratio than the iodine observations and using a narrower slit for higher resolution. Since the PSF is constantly changing on a slit-fed spectrograph due to changes in the slit illumination caused by variable seeing and guiding patterns, we need to deconvolve this stellar template to remove distortions in the line profile. This deconvolution process also requires a measurement of the instrumental PSF. By construction, we do not have a simultaneous PSF calibration for the template observations as we do for the iodine observations. We rely on bracketing observations of rapidly rotating B stars through iodine, but the PSF is never exactly the same for any two observations and this non-simultaneous calibration introduces errors. In addition, the deconvolution process is inherently unstable because we are essentially trying to create a view of the star at infinite spectral resolution using data taken at finite spectral resolution. Collecting as much signal as possible for the template observation helps ensure that this component of the model is not the limiting factor on the final RV precision.

For faint stars (V$>$12), iodine exposures on Keck/HIRES can take 45-60 min each. Collecting a template at 2-3 times higher S/N ratio would take a significant fraction of a night. At APF the situation is even worse due to its much smaller aperture (2.4 m vs. 10 m). When collecting reconnaissance RVs for a new candidate from a transit survey to look for various false positive scenarios only 2-3 RVs may be needed before a false alarm scenario can be confirmed or eliminated. Collecting a high S/N ratio template can take as much time as the entire set of RV observations, effectively doubling the observational time needed in these scenarios.

In order to circumvent these problems we developed a technique that uses a synthetic template using the \citet{Coelho14} models and spectral synthesis code from SpecMatch. We first determine the spectral parameters from the best available information. Approximate values for $\teff$, $\logg$, and $\feh$ can be obtained using spectral regions outside of the iodine absorption regime, photometry, or low resolution spectroscopy. We normally use a single echelle order redward of the iodine regime to estimate $\vsini$. $\teff$ and $\vsini$ have the most drastic impact on the resulting RV, but fortunately \teff\ is the easiest parameter to estimate in the absence of a high resolution spectrum and $\vsini$ can be estimated with very few spectral lines. Once the spectral parameters are established we synthesize the full spectrum within the iodine region setting the instrumental PSF width to zero. This provides an extremely high-resolution, noise-free stellar template to use as one of the inputs to the normal RV extraction code. This foregoes the need for deconvolution and saves valuable telescope time.

\begin{figure}[htbp] 
   \centering
\includegraphics[width=3.5in]{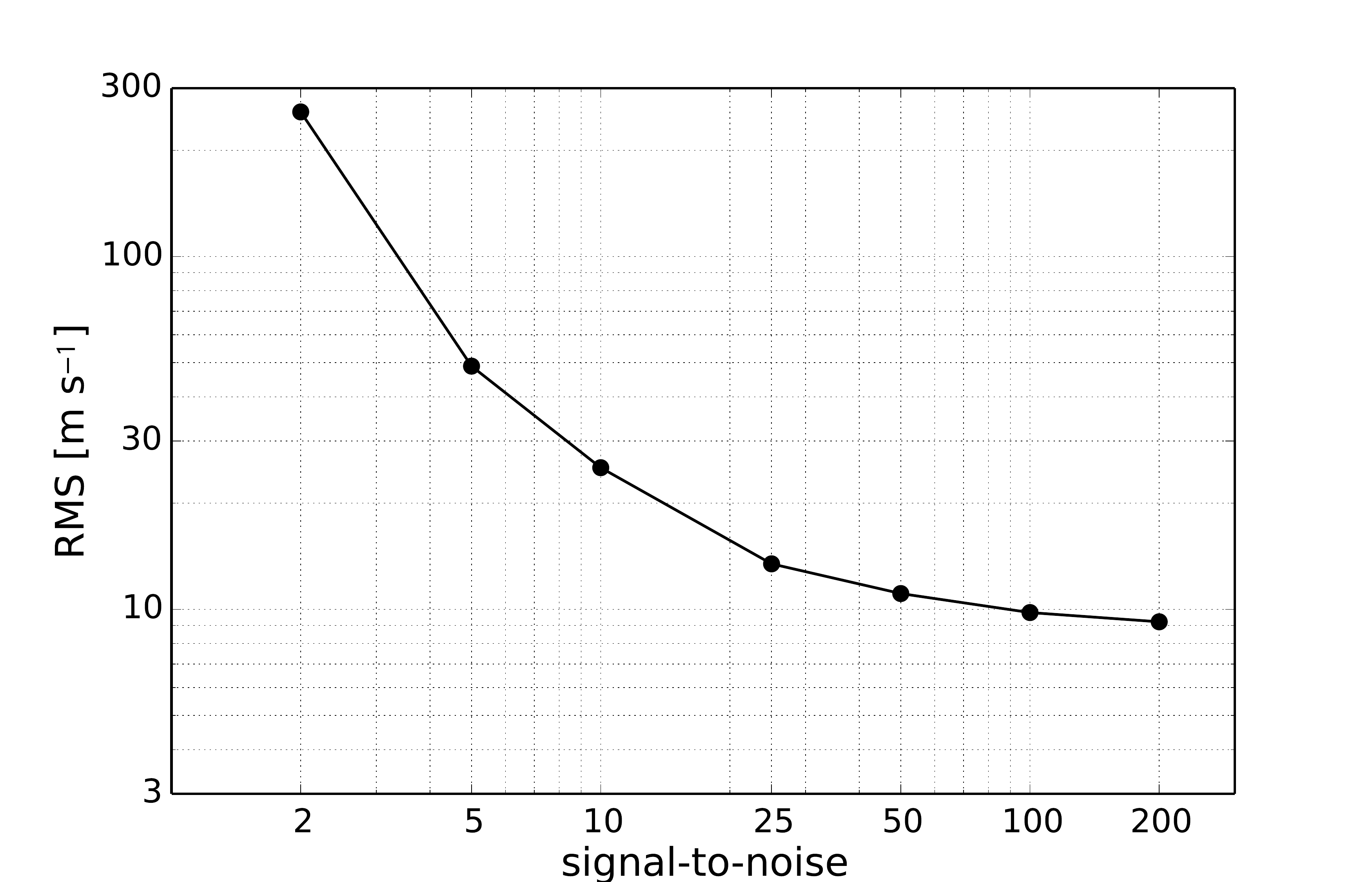}
   \caption{Synthetic template RV performance as a function of S/N ratio of the iodine observations. Artificial Gaussian noise was injected into the spectra of a well-known RV standard star (HD 9407) before running them through the RV extraction pipeline for all data points excluding the one at S/N=200. The synthetic template technique shows predictable and consistent performance down to extremely low S/N ratios and a model-limited noise floor of $\approx$10 \mse.}
   \label{fig:synthRV}
\end{figure}

\begin{figure}[htbp] 
   \centering
\includegraphics[width=3.5in]{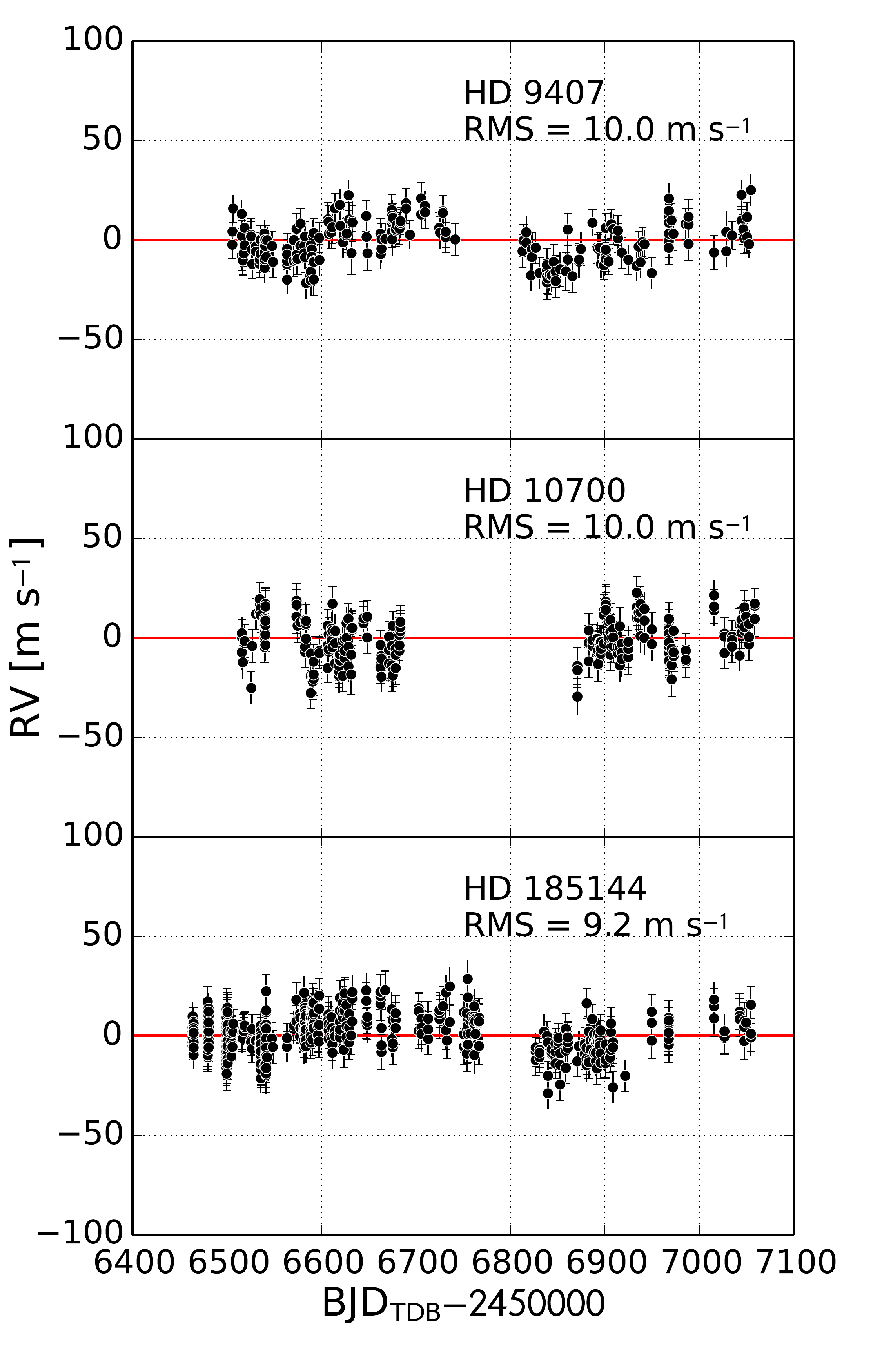}
   \caption{Synthetic template RV performance for three RV standard stars. Artificial Gaussian noise was injected into all spectra for each star so that S/N=50 for each spectrum. This is approximately the same S/N ratio that was collected for each KELT-8 observation. The RV RMS is about 10 \ms for each star.}
   \label{fig:synth_standards}
\end{figure}

This technique is not perfect due to inaccuracies in the model spectra. Spectral lines are missing in some regions, blended lines may only be accounted for as a single line, or the central wavelengths of lines may be slightly inaccurate. This means that using these synthetic templates we can never achieve the 2-3 \ms precision that we can obtain using the traditional technique for bright stars. Figure \ref{fig:synthRV} shows velocity RMS for the RV standard star (HD 9407) as a function of S/N ratio of the iodine observations using the synthetic template technique. We artificially inject Gaussian noise into the spectra before sending them through the normal RV extraction pipeline. We use an exposure meter to expose to S/N=200 for all normal observations and the highest S/N ratio datapoint shown in Figure \ref{fig:synthRV} has no artificial noise injected. The RV RMS monotonically decreases down to extremely low S/N ratio levels with a noise floor at just below 10 \ms for high S/N ratio. Two other RV standard stars were also run through the synthetic template pipeline with S/N ratio artificially decreased to match that of our KELT-8 observations. In each case the synthetic template RV RMS was about 10 \ms (see Figure \ref{fig:synth_standards}). We use this technique to extract the RVs from the APF data for KELT-8 and achieve a median per-measurement precision of 11.8 \mse.

While our technique is unique for iodine-based RVs in the optical, several other very similar techniques have previously been developed to extract high precision RVs. \citet{Bailey12} and \citet{Tanner12} use interpolated model atmospheres for their templates, but instead of an iodine cell they use telluric lines for their PSF and wavelength calibration. They were able to achieve $\sim$50 m/s precision in the near infrared with NIRSPEC on Keck. Other telluric-calibrated techniques have proven successful in the optical with HIRES to obtain velocities to ~100 m/s precision \citep{Chubak12}. \citet{Johnson06} perturb an existing empirical template of a similar star in order to avoid the need to collect a unique template for each star observed. They were able to achieve a prescision of 3-5 \ms using these perturbed empirical templates.

\section{Planetary Characterization}
\label{sec:characterization}
\subsection{Global Fit}

We determine the system's orbital and planetary properties by performing a global analysis on the RVs, spectroscopic stellar parameters, and follow-up light curves using a custom version of EXOFAST \citep{Eastman13}. Our global analysis technique is very similar to that used for the previous KELT discoveries \citep{Siverd12, Beatty12, Pepper13, Collins14, Bieryla15}. The EXOFAST package uses a DE-MCMC fitting routing to extract the median parameter values and 68\% confidence intervals using all available data and these values are presented in Table \ref{tab:params}. Constraints on the stellar parameters $M_*$ and $R_*$ are also included using the Yonsei-Yale stellar models and spectroscopically derived values for \teff\ and \feh\ as input priors. Before running the MCMC chains, a best-fit solution is found using an AMOEBA $\chi^2$ minimization routine and the photometric and RV uncertainties are scaled such that \chisq=1. The scaling factors for the photometric data are listed in Table \ref{tab:photobs} and the scaling factor applied to the APF RV uncertainties is 1.44. The scaled uncertainties include both instrumental and astrophysical noise sources and this ensures that the widths of final posterior properly represent the uncertainties in the model parameters.

As described in Collins et al. (2014), we identify trend datasets that best improve the individual light curve fits using AIJ. These trend datasets are included as input to EXOFAST, along with the undetrended light curve data.  As part of the global fit, EXOFAST simultaneously finds the best transit model fit and removes linear trends that are correlated with different combinations of airmass, time, FWHM, and CCD position.

The light from the visual companion to KELT-8 was included in the photometric aperture of the GCorfini, and ZRO light curves. We corrected for this by measuring the flux ratio of the primary to secondary component in the well resolved MORC data. The flux ratios are 1.24 and 1.44 in $g'$ and $i'$ band respectively. We then interpolated/extrapolated these flux ratios to the central wavelengths of the V and Ic GCorfini and ZRO light curves and corrected the input light curves for the dilution before fitting them in EXOFAST. We also ran the global fit excluding these two transits and found that all parameters are consistent to well within 1 $\sigma$. This gives us confidence that these blended light curves are not biasing the inferred parameters.

We ran four permutations of the global fit using slightly different input priors and constraints.  First, we assert constraints on the stellar mass and radius from the Yonsei-Yale isochrones and a uniform prior on \logg. We adopt the parameter values from this fit for all analysis and interpretation. Second, we performed the same fit using constraints from the \citet{Torres10} relations instead. These two fits produced nearly identical results. Then we performed these same two global fits but imposed a prior on \logg\ from the high-resolution spectroscopy. The \logg\ derived from spectroscopy is slightly higher and only marginally consistent with the \logg\ derived from transits. The results from all four fits are presented in Table \ref{tab:params}. The differences in parameter values from these four fits give an estimate of our systematic uncertainty which appears to be similar to the statistical uncertainties for most parameters. This gives us confidence that our parameter uncertainties are properly estimated and that our interpretations of physical parameters (such as the large planetary radius) are robust.

We measure an eccentricity for KELT-8b of $e=0.035_{-0.025}^{+0.050}$ which deviates from zero by only 1.4 $\sigma$. The measurement of eccentricity is easily biased to artificially larger values and a significance of $\geq$2.5 $\sigma$ is generally accepted as being required to claim a non-circular orbit \citep{Lucy71}. Although we do not find any significant eccentricity in the system, moderate eccentricity is not unprecedented in other hot Jupiters \citep[e.g. HAT-P-2b, XO-3b, WASP14b, HAT-P-31b][]{Bakos07, Winn09, Joshi09, Kipping11}. We allow for eccentricity to vary as a free parameter in order to ensure that the errors on all other parameters are not underestimated. 

\begin{figure}[h] 
   \centering
\includegraphics[width=3.5in]{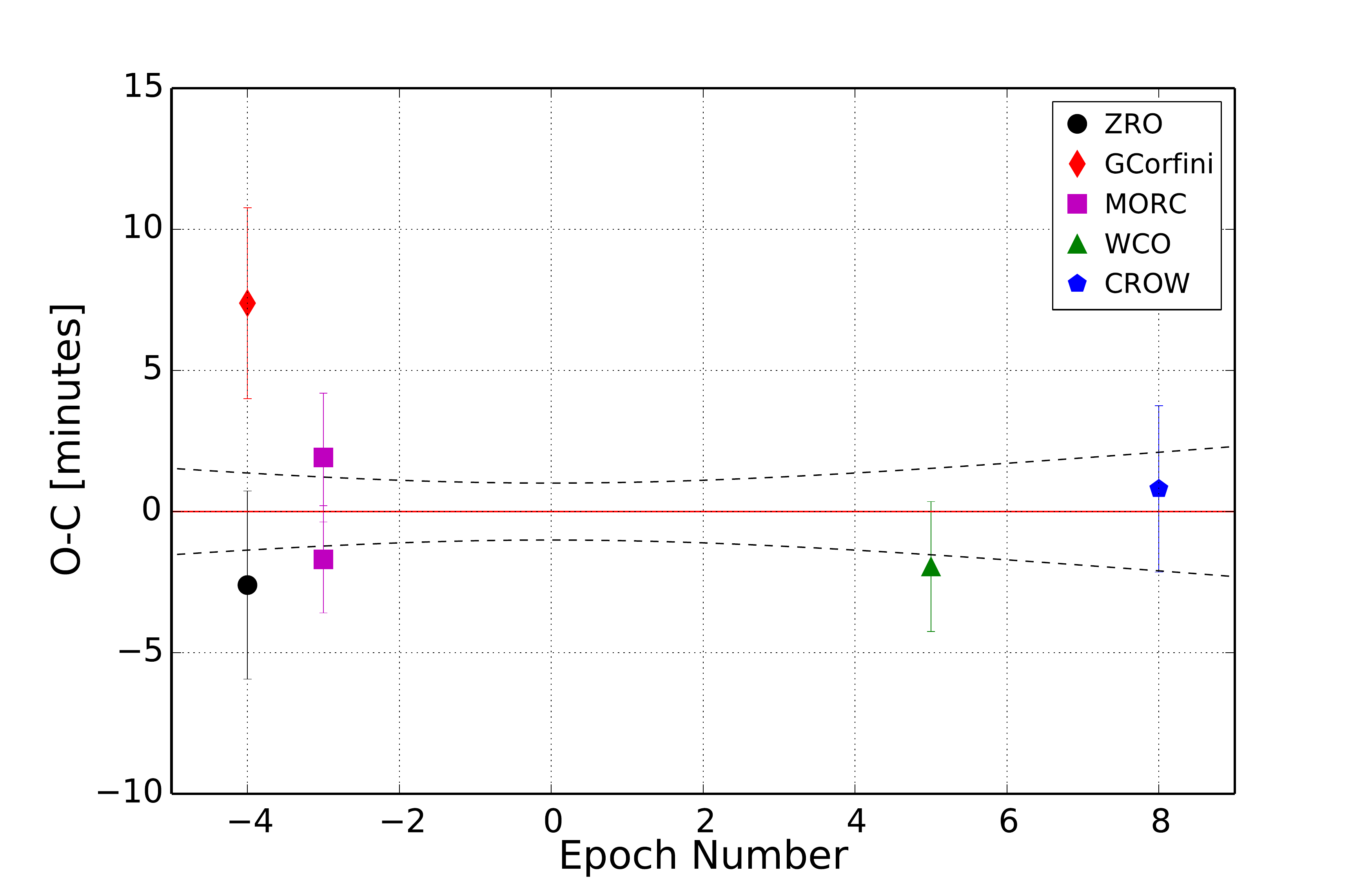}
   \caption{Residuals of the transit times from the best-fit ephemeris for the followup light curves. The source of each transit time is denoted in the upper right. The dashed lines indicate the 1 $\sigma$ uncertainty on the linear ephemeris.}
   \label{fig:ttv}
\end{figure}

We searched for possible transit timing variations (TTVs) in the system by allowing the transit times for each of the follow-up light curves to vary. The ephemeris is constrained by the RV data and a prior imposed from the KELT discovery data. We find no evidence for significant transit timing variations. The individual transit times are presented in Table \ref{tab:ttv} and Figure \ref{fig:ttv}.

After the completion of the global fit we then derive a linear ephemeris. The period is constrained by the RV data and comes from the global fit. We fit for T$_0$ using the transit times from the followup light curves. The best-fit linear ephemeris is listed in Table \ref{tab:params}. We find $P = 3.24406 \pm 0.00016$ days and $T_0 = 2456883.4803 \pm 0.0007$ \bjdtdb. The $\chi^2$ for the linear ephemeris fit is 7.4 with 4 degrees of freedom which indicates that a linear ephemeris adequately describes the observed transit times.

\begin{figure}[h]
   \centering
\includegraphics[width=0.48\textwidth]{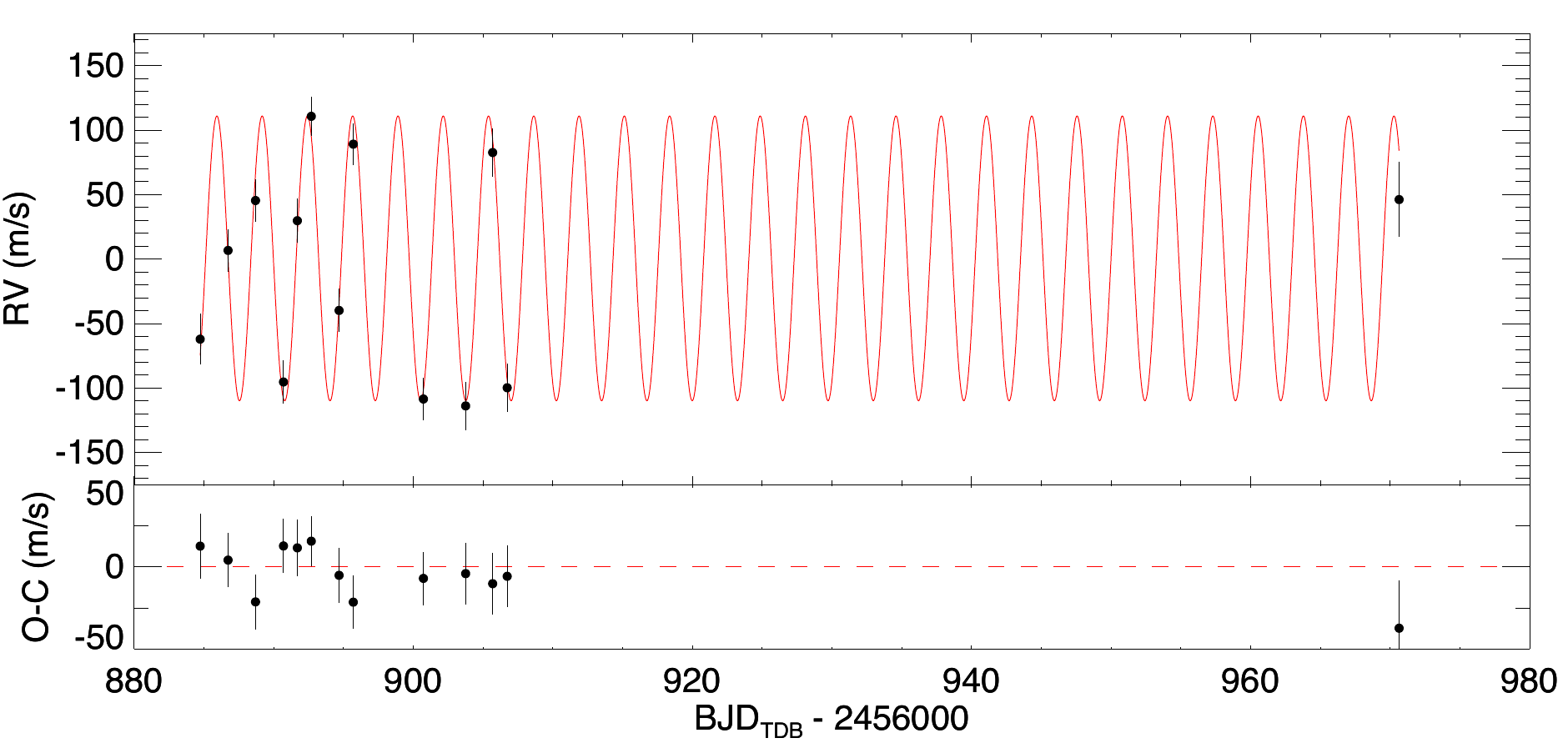}
\includegraphics[width=0.48\textwidth]{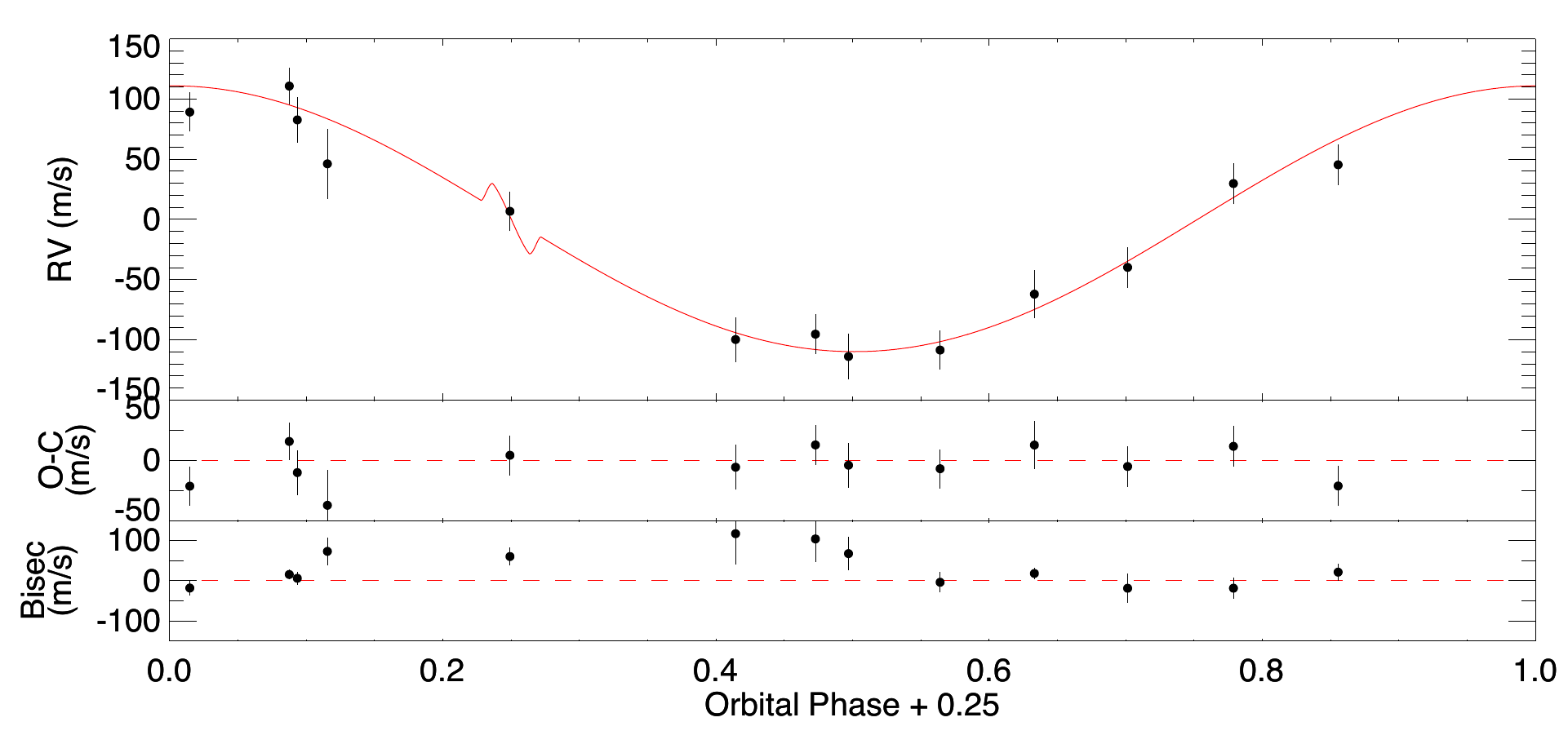}
   \caption{\emph{Top:} APF RVs and residuals for KELT-8. The best-fit model is shown in red. \emph{Bottom:} RVs and bisector span measurements phase-folded to the best-fit linear ephemeris. The best-fit model is shown in red. The predicted Rossiter-McLaughlin effect assumes perfect spin-orbit alignment and it is not constrained by our data.}
   \label{fig:RV}
\end{figure}

\begin{deluxetable*}{lccccc}
\tablecaption{Median values and 68\% confidence intervals for KELT-8b.}
\tablehead{\colhead{~~~Parameter} & \colhead{Units} & \colhead{{\bf Adopted Value}}	&  \colhead{Value}	&  \colhead{Value}	&  \colhead{Value} \\
\colhead{} & \colhead{} & \colhead{{\bf (YY)}}	&  \colhead{(Torres)}	&  \colhead{(YY+\logg\ prior)}	&  \colhead{(Torres+\logg\ prior)}}
\startdata
\sidehead{Stellar Parameters:}	
                               ~~~$M_{*}$\dotfill &Mass (\msun)\dotfill & $1.211_{-0.066}^{+0.078}$	&  $1.233_{-0.063}^{+0.066}$	&  $1.188_{-0.063}^{+0.067}$	&  $1.211_{-0.059}^{+0.060}$\\
                             ~~~$R_{*}$\dotfill &Radius (\rsun)\dotfill & $1.67_{-0.12}^{+0.14}$	&  $1.69_{-0.12}^{+0.15}$	&  $1.570_{-0.097}^{+0.099}$	&  $1.577\pm0.096$\\
                         ~~~$L_{*}$\dotfill &Luminosity (\lsun)\dotfill & $2.74_{-0.40}^{+0.51}$	&  $2.79_{-0.41}^{+0.54}$	&  $2.42_{-0.32}^{+0.35}$	&  $2.44_{-0.31}^{+0.34}$\\
                             ~~~$\rho_*$\dotfill &Density (cgs)\dotfill & $0.369_{-0.067}^{+0.073}$	&  $0.363_{-0.072}^{+0.078}$	&  $0.433_{-0.061}^{+0.076}$	&  $0.436_{-0.062}^{+0.078}$\\
                  ~~~$\log{g_*}$\dotfill &Surface gravity (cgs)\dotfill & $4.078_{-0.054}^{+0.049}$	&  $4.075_{-0.061}^{+0.054}$	&  $4.121_{-0.041}^{+0.043}$	&  $4.125_{-0.042}^{+0.045}$\\
                  ~~~$\teff$\dotfill &Effective temperature (K)\dotfill & $5754_{-55}^{+54}$	&  $5749\pm59$	&  $5748\pm53$	&  $5750\pm59$\\
                                 ~~~$\feh$\dotfill &Metallicity\dotfill & $0.272\pm0.038$	&  $0.270\pm0.039$	&  $0.270_{-0.039}^{+0.038}$	&  $0.271_{-0.039}^{+0.040}$\\
\sidehead{Planetary Parameters:}
                                   ~~~$e$\dotfill &Eccentricity\dotfill & $0.035_{-0.025}^{+0.050}$	&  $0.037_{-0.027}^{+0.057}$	&  $0.028_{-0.019}^{+0.033}$	&  $0.027_{-0.020}^{+0.033}$\\
        ~~~$\omega_*$\dotfill &Argument of periastron (degrees)\dotfill & $85_{-97}^{+87}$	&  $85_{-93}^{+82}$	&  $-80_{-130}^{+120}$	&  $-80_{-130}^{+120}$\\
                           ~~~$a$\dotfill &Semi-major axis (AU)\dotfill & $0.04571_{-0.00084}^{+0.00096}$	&  $0.04599_{-0.00079}^{+0.00081}$	&  $0.04542_{-0.00082}^{+0.00084}$	&  $0.04571_{-0.00075}^{+0.00074}$\\
                                 ~~~$M_{P}$\dotfill &Mass (\mj)\dotfill & $0.867_{-0.061}^{+0.065}$	&  $0.876_{-0.061}^{+0.063}$	&  $0.855_{-0.061}^{+0.062}$	&  $0.869_{-0.060}^{+0.061}$\\
                               ~~~$R_{P}$\dotfill &Radius (\rj)\dotfill & $1.86_{-0.16}^{+0.18}$	&  $1.88_{-0.16}^{+0.19}$	&  $1.73\pm0.13$	&  $1.74\pm0.13$\\
                           ~~~$\rho_{P}$\dotfill &Density (cgs)\dotfill & $0.167_{-0.038}^{+0.047}$	&  $0.163_{-0.039}^{+0.049}$	&  $0.205_{-0.039}^{+0.051}$	&  $0.205_{-0.040}^{+0.052}$\\
                      ~~~$\log{g_{P}}$\dotfill &Surface gravity\dotfill & $2.793_{-0.075}^{+0.072}$	&  $2.786_{-0.080}^{+0.076}$	&  $2.850_{-0.063}^{+0.065}$	&  $2.853_{-0.064}^{+0.066}$\\
               ~~~$T_{eq}$\dotfill &Equilibrium temperature (K)\dotfill & $1675_{-55}^{+61}$	&  $1679_{-57}^{+66}$	&  $1629_{-48}^{+47}$	&  $1628_{-47}^{+46}$\\
                           ~~~$\Theta$\dotfill &Safronov number\dotfill & $0.0351_{-0.0037}^{+0.0040}$	&  $0.0346_{-0.0038}^{+0.0041}$	&  $0.0378_{-0.0035}^{+0.0039}$	&  $0.0377_{-0.0035}^{+0.0039}$\\
                   ~~~$\fave$\dotfill &Incident flux (\fluxcgs)\dotfill & $1.78_{-0.22}^{+0.27}$	&  $1.80_{-0.23}^{+0.29}$	&  $1.60_{-0.18}^{+0.19}$	&  $1.59_{-0.18}^{+0.19}$\\
\sidehead{RV Parameters:}
               ~~~$T_{P}$\dotfill &Time of periastron (\bjdtdb-2450000)\dotfill & $6870.47_{-0.86}^{+0.75}$	&  $6870.47_{-0.83}^{+0.70}$	&  $6869.0_{-1.2}^{+1.1}$	&  $6869.0_{-1.2}^{+1.1}$\\
                        ~~~$K$\dotfill &RV semi-amplitude (m/s)\dotfill & $104.0\pm6.4$	&  $104.0\pm6.4$	&  $104.1\pm6.5$	&  $104.4\pm6.5$\\
                           ~~~$M_{P}/M_{*}$\dotfill &Mass ratio\dotfill & $0.000683_{-0.000044}^{+0.000045}$	&  $0.000679_{-0.000043}^{+0.000044}$	&  $0.000687_{-0.000044}^{+0.000045}$	&  $0.000685\pm0.000044$\\
                                 ~~~$\gamma_{APF}$\dotfill &m/s\dotfill & $6.1\pm4.9$	&  $6.0\pm5.0$	&  $6.7\pm4.9$	&  $6.7_{-5.0}^{+4.9}$\\
                                         ~~~$\ecosw$\dotfill & \dotfill & $0.001_{-0.021}^{+0.024}$	&  $0.001_{-0.021}^{+0.025}$	&  $0.001_{-0.019}^{+0.022}$	&  $0.001_{-0.018}^{+0.022}$\\
                                         ~~~$\esinw$\dotfill & \dotfill & $0.012_{-0.025}^{+0.063}$	&  $0.014_{-0.027}^{+0.071}$	&  $-0.001_{-0.032}^{+0.029}$	&  $-0.001_{-0.034}^{+0.028}$\\
\sidehead{Primary Transit Parameters:}
~~~$R_{P}/R_{*}$\dotfill &Planet to star radius ratio\dotfill & $0.1145\pm0.0026$	&  $0.1146\pm0.0026$	&  $0.1133\pm0.0025$	&  $0.1132\pm0.0025$\\
           ~~~$a/R_*$\dotfill &Semi-major axis in stellar radii\dotfill & $5.90_{-0.38}^{+0.37}$	&  $5.87_{-0.42}^{+0.39}$	&  $6.22_{-0.31}^{+0.34}$	&  $6.23_{-0.31}^{+0.35}$\\
                          ~~~$i'$\dotfill &Inclination (degrees)\dotfill & $82.65_{-1.0}^{+0.81}$	&  $82.56_{-1.2}^{+0.88}$	&  $83.36\pm0.69$	&  $83.38_{-0.70}^{+0.69}$\\
                               ~~~$b$\dotfill &Impact parameter\dotfill & $0.741_{-0.033}^{+0.027}$	&  $0.743_{-0.033}^{+0.027}$	&  $0.722_{-0.036}^{+0.028}$	&  $0.722_{-0.037}^{+0.029}$\\
                             ~~~$\delta$\dotfill &Transit depth\dotfill & $0.01311\pm0.00059$	&  $0.01313_{-0.00059}^{+0.00061}$	&  $0.01283_{-0.00057}^{+0.00058}$	&  $0.01282\pm0.00057$\\
              ~~~$\tau$\dotfill &Ingress/egress duration (days)\dotfill & $0.0303_{-0.0035}^{+0.0038}$	&  $0.0305_{-0.0036}^{+0.0039}$	&  $0.0280_{-0.0032}^{+0.0033}$	&  $0.0280_{-0.0032}^{+0.0034}$\\
                     ~~~$T_{14}$\dotfill &Total duration (days)\dotfill & $0.1444_{-0.0033}^{+0.0034}$	&  $0.1446\pm0.0034$	&  $0.1423_{-0.0031}^{+0.0032}$	&  $0.1423\pm0.0032$\\
                     ~~~$u_{1I}$\dotfill &Linear Limb-darkening\dotfill & $0.2961_{-0.0088}^{+0.0092}$	&  $0.2966_{-0.0093}^{+0.0096}$	&  $0.2977_{-0.0087}^{+0.0089}$	&  $0.2975_{-0.0092}^{+0.0096}$\\
                  ~~~$u_{2I}$\dotfill &Quadratic Limb-darkening\dotfill & $0.2793_{-0.0049}^{+0.0044}$	&  $0.2790_{-0.0052}^{+0.0047}$	&  $0.2782_{-0.0047}^{+0.0042}$	&  $0.2783_{-0.0051}^{+0.0046}$\\
                ~~~$u_{1Sloang'}$\dotfill &Linear Limb-darkening\dotfill & $0.607\pm0.015$	&  $0.607_{-0.015}^{+0.016}$	&  $0.608\pm0.015$	&  $0.608_{-0.015}^{+0.016}$\\
             ~~~$u_{2Sloang'}$\dotfill &Quadratic Limb-darkening\dotfill & $0.184_{-0.011}^{+0.010}$	&  $0.184_{-0.012}^{+0.011}$	&  $0.183_{-0.011}^{+0.010}$	&  $0.184_{-0.012}^{+0.011}$\\
                ~~~$u_{1Sloani'}$\dotfill &Linear Limb-darkening\dotfill & $0.3178_{-0.0092}^{+0.0097}$	&  $0.3183_{-0.0097}^{+0.010}$	&  $0.3193_{-0.0091}^{+0.0094}$	&  $0.3192_{-0.0096}^{+0.010}$\\
             ~~~$u_{2Sloani'}$\dotfill &Quadratic Limb-darkening\dotfill & $0.2789_{-0.0052}^{+0.0046}$	&  $0.2786_{-0.0056}^{+0.0050}$	&  $0.2779_{-0.0051}^{+0.0045}$	&  $0.2779_{-0.0055}^{+0.0049}$\\
                ~~~$u_{1Sloanr'}$\dotfill &Linear Limb-darkening\dotfill & $0.410_{-0.011}^{+0.012}$	&  $0.410\pm0.012$	&  $0.411_{-0.011}^{+0.012}$	&  $0.411\pm0.012$\\
             ~~~$u_{2Sloanr'}$\dotfill &Quadratic Limb-darkening\dotfill & $0.2725_{-0.0070}^{+0.0062}$	&  $0.2722_{-0.0074}^{+0.0066}$	&  $0.2715_{-0.0068}^{+0.0062}$	&  $0.2717_{-0.0074}^{+0.0066}$\\
                     ~~~$u_{1V}$\dotfill &Linear Limb-darkening\dotfill & $0.487_{-0.012}^{+0.013}$	&  $0.488_{-0.013}^{+0.014}$	&  $0.488_{-0.012}^{+0.013}$	&  $0.488_{-0.013}^{+0.014}$\\
                  ~~~$u_{2V}$\dotfill &Quadratic Limb-darkening\dotfill & $0.2446_{-0.0088}^{+0.0077}$	&  $0.2442_{-0.0093}^{+0.0082}$	&  $0.2435_{-0.0087}^{+0.0078}$	&  $0.2437_{-0.0093}^{+0.0082}$\\
                  
\sidehead{Linear Ephemeris from Follow-up Transits:}
	~~~$P$\dotfill &Period (days)\dotfill & $3.24406\pm0.00016$  &   $3.24406\pm0.00016$   &  $3.24406\pm0.00016$  &  $3.24406\pm0.00016$\\
	~~~$T_0$\dotfill &Time of transit (\bjdtdb-2450000)\dotfill & $6883.4803\pm0.0007$   &  $6883.4803\pm0.0007$  &   $6883.4804\pm0.0007$  & $6883.4804\pm0.0007$ \\
	
\sidehead{Predicted Secondary Eclipse Parameters:}
                  ~~~$T_{S}$\dotfill &Time of eclipse (\bjdtdb-2450000)\dotfill & $6868.886_{-0.043}^{+0.049}$	&  $6868.887_{-0.044}^{+0.050}$	&  $6868.886_{-0.039}^{+0.044}$	&  $6872.130_{-0.038}^{+0.044}$\\
                           ~~~$b_{S}$\dotfill &Impact parameter\dotfill & $0.763_{-0.057}^{+0.097}$	&  $0.769_{-0.061}^{+0.11}$	&  $0.718_{-0.054}^{+0.051}$	&  $0.716_{-0.055}^{+0.051}$\\
            ~~~$\tau_S$\dotfill &Ingress/egress duration (days)\dotfill & $0.0327_{-0.0059}^{+0.016}$	&  $0.0332_{-0.0065}^{+0.019}$	&  $0.0277_{-0.0042}^{+0.0054}$	&  $0.0276_{-0.0042}^{+0.0053}$\\
                   ~~~$T_{S,14}$\dotfill &Total duration (days)\dotfill & $0.1436_{-0.0049}^{+0.0034}$	&  $0.1435_{-0.0060}^{+0.0036}$	&  $0.1419_{-0.0041}^{+0.0033}$	&  $0.1418_{-0.0042}^{+0.0034}$\\
\enddata
\label{tab:params}
\end{deluxetable*}

\begin{deluxetable}{lccccl}
\tablecaption{Transit Times for KELT-8b.}
\tablehead{\colhead{Epoch} & \colhead{$T_C$} & \colhead{$\sigma_{T_C}$} & \colhead{O$-$C} & \colhead{O$-$C} & \colhead{Telescope}\\
		\colhead{} & \colhead{(\bjdtdb)} & \colhead{(s)} & \colhead{(s)} & \colhead{($\sigma_{T_C}$)} & \colhead{}}
\startdata
-4 & 2456870.502251 & 200  &  -156.3 & -0.78  &    ZRO  \\
-4 & 2456870.509188 & 203  & 443.1 & 2.18 &	  GCorfini \\
-3 & 2456873.746945 & 114  &  -101.5  & -0.89  &   MORC  \\
-3 & 2456873.749450 & 137   & 114.9 & 0.83  &   MORC  \\
5 & 2456899.699249 & 138   &  -116.7 & -0.85 &   WCO \\
8 & 2456909.433338 & 177  &  48.2 & 0.27 &   CROW  \\

\enddata
\label{tab:ttv}
\end{deluxetable}

\subsection{False-Positive Assessment}
\label{sec:fap}

We calculate bisector spans (BS) for each of the APF spectra in order to investigate possible blend scenarios in which the measured RV variations are caused by line distortions that arise from an unresolved eclipsing binary instead of true reflex motion of the star. In the case that the measured Doppler shifts are caused by a second set of incompletely resolved stellar lines from a luminous companion we would expect to see BSs that vary in phase with the orbital signal and a correlation in BS vs. RV \citep{Santos02, Torres05}. The amplitude would be similar to that of the velocities \citep{Queloz01, Manushev05}.

Following the prescription of \citet{Torres05} and \citet{Torres07}, each APF spectrum was cross-correlated with a synthetic spectrum derived using the stellar properties determined from the high S/N ratio Keck spectrum (see Section \ref{sec:stellar_props}). We analyzed the cross-correlation function for each of the echelle orders between 4260 and 5000 \AA\ in order to avoid the iodine lines, and telluric lines. By restricting the BS analysis to blue orders we reduce the effect of instrumental PSF variations caused by guiding errors because the seeing is degraded toward the blue which means that the slit is more evenly illuminated. We measure the asymmetry in the spectral line profile for each order by calculating the velocity at the midpoint of lines connecting the CCF at many different fractional levels of CCF peak. The BS is then the difference in velocity between the 65th and 95th percentile levels of the CCF. The bisector spans and error corresponding to a given observation are the mean and standard deviation on the mean over the 15 spectral orders analyzed.

\begin{figure}[h] 
   \centering
\includegraphics[width=3.5in]{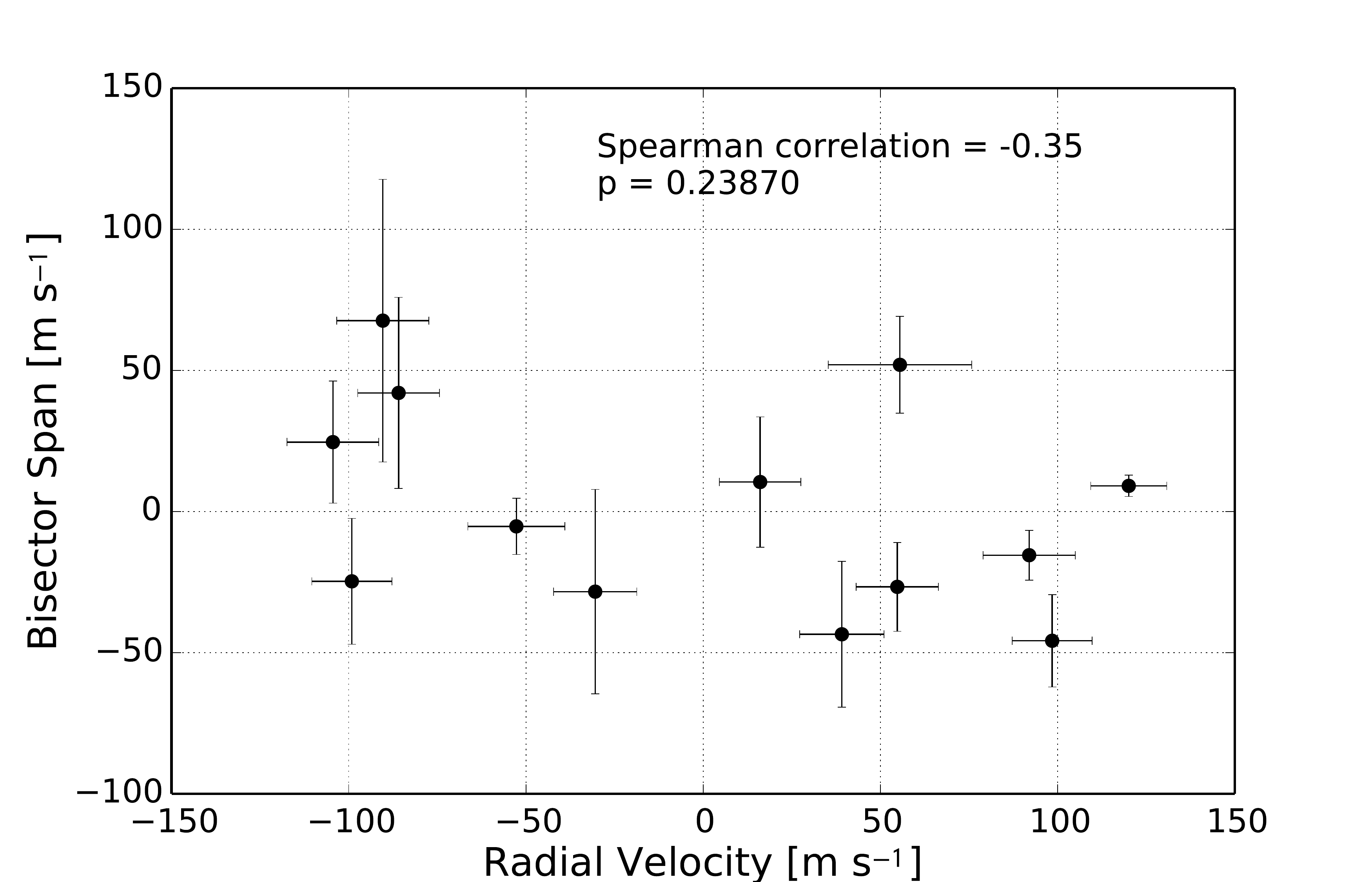}
   \caption{BS measurements for the APF spectra used for radial velocity measurements. We find no significant correlation between RV and BS.}
   \label{fig:BS}
\end{figure}

Our BS measurements and errors are listed in Table \ref{tab:rv} and shown in Figure \ref{fig:BS}. Figure \ref{fig:RV} also shows the BS measurements as a function of orbital phase. We find no statistically significant correlation of BS with RV. The Spearman rank correlation coefficient is -0.35 \citep[p=0.24,][]{Spearman1904}.

The lack of in-phase BS variations, 
and the fact that the stellar \logg\ derived from spectroscopy and \logg\ derived from the transit light curves are consistent to within 1-$\sigma$ lead us to conclude that the RV variations and transit signals are caused by a highly-inflated Jovian planet. 

\section{Discussion}
\label{sec:discussion}


\subsection{Prospects for further characterization of the host star}
Further stellar characterization of KELT-8 is needed to better constrain the mass and radius of the star which will improve our knowledge of the planetary mass and radius. Several more light curves with higher precision in which the two components are well-resolved would help to better constrain the stellar density. A more detailed analysis of higher S/N ratio spectra may help to nail down the spectroscopically derived parameters and allow for chemical abundance measurements. GAIA \citep{deBruijne12} should provide a precise parallax to constrain the stellar luminosity and thus radius, and TESS and/or PLATO \citep{Rauer14} may provide sufficient photometric precision to measure the stellar density using asteroseismology.

\subsection{Comparative Planetology}
\begin{figure}[t]
   \centering
\includegraphics[width=3.5in]{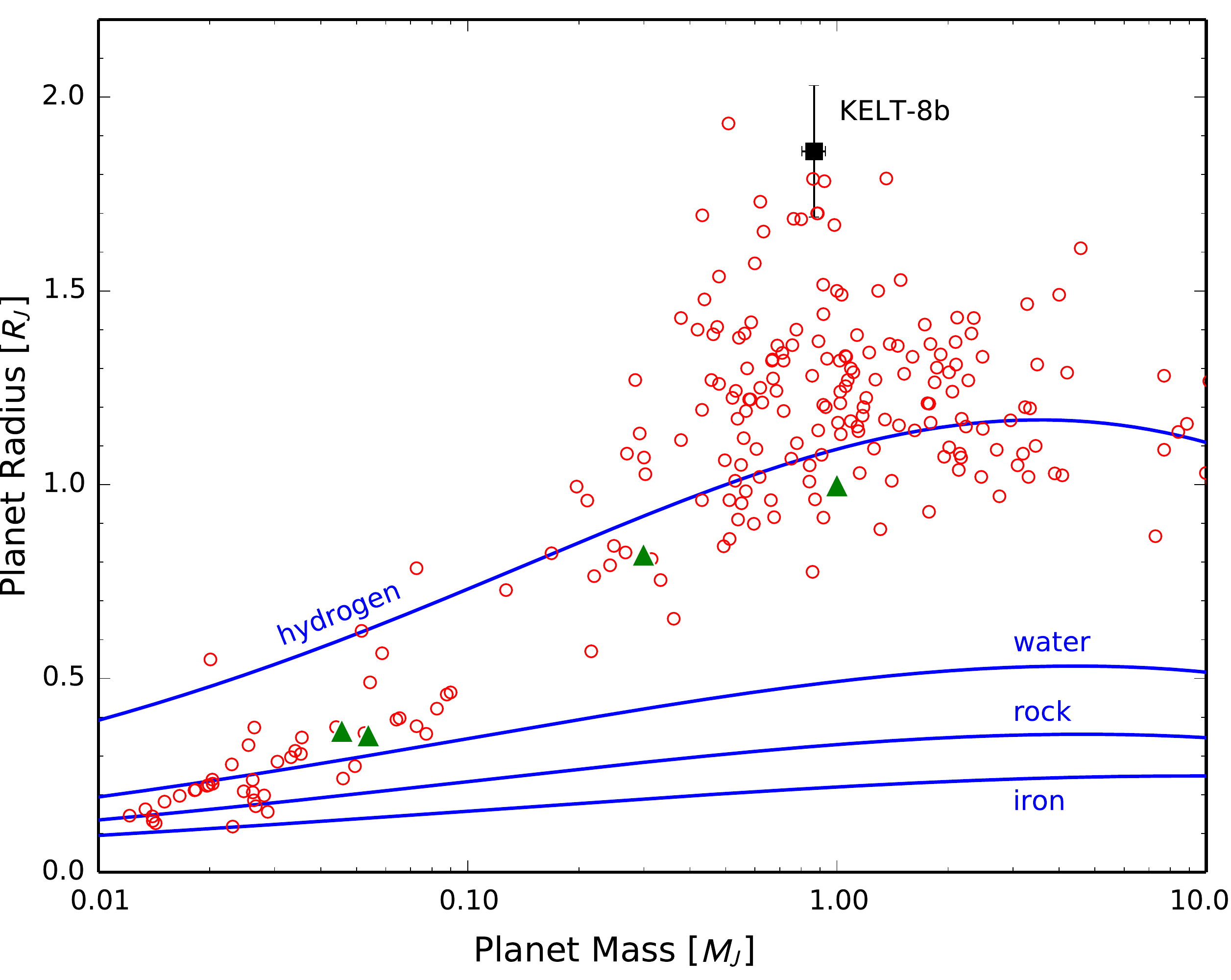}
   \caption{Mass radius diagram of all confirmed transiting exoplanets\footnote{based on a 2/17/2015 query of exoplanets.org} (red circles). KELT-8b is the annotated black square. Solar system planets are represented by green triangles. Model mass-radius relationships for idealized planets consisting of pure hydrogen, water, rock (Mg$_2$SiO$_4$), and iron are shown as blue lines \citep{Seager07}. Only WASP-17b has a radius larger than the best-fit radius for KELT-8b.}
   \label{fig:mr}
\end{figure}

KELT-8b has the 2nd largest radius of all known exoplanets, but has a mass slightly less than that of Jupiter. It lies well above the theoretical mass-radius curve for pure hydrogen (see Figure \ref{fig:mr}). KELT-8b is similar to WASP-1b \citep{Collier07}, and HAT-P-41b \citep{Hartman12}.
It orbits a star more massive and metal rich than the Sun (\mstar $\geq1$ \msun and \feh $\geq0.1$), the planet is highly irradiated ($a \leq 0.05$ AU), and it is less massive than Jupiter but extremely inflated ($R_p \geq 1.5$ \rj).

\begin{figure}[t]
   \centering
\includegraphics[width=3.5in]{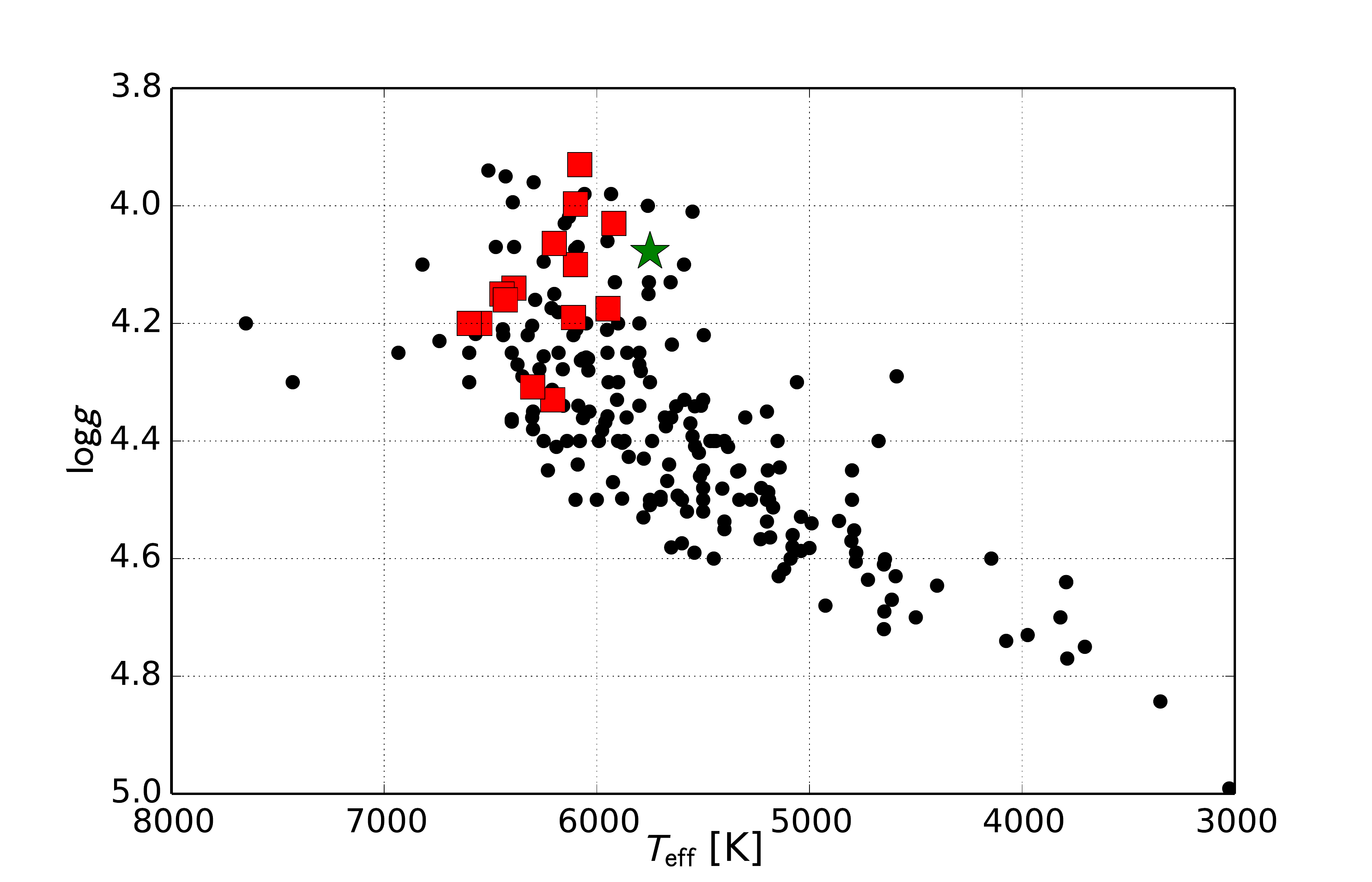}
   \caption{HR diagram showing well-characterized host stars of hot Jupiters. The red squares indicate stars that host planets with radii greater than 1.5 \rj. The green star marks the position of KELT-8, and the black circles are all other systems.}
   \label{fig:HR}
\end{figure}

Next, we compare the stellar parameters of KELT-8 with other well-characterized hot Jupiter host stars. We use the exoplanets.org catalogue to select the stars that host planets with orbital periods shorter than 10 days, and masses between 0.4 and 1.1 \mj. We plot these stars on an HR diagram in Figure \ref{fig:HR}. We split this sample into two groups; one group containing the stars that host planets with radii greater than 1.5 \rj\, and the other group containing stars that host planets smaller than 1.5 \rj.
KELT-8 is the coolest star to host a highly inflated hot Jupiter ($\teff=5754$ K), but we find that all other stars that host similar highly inflated planets tend to be hotter than $\approx$5900 K. However, there are also plenty of stars hotter than 5900 K that host non-inflated hot Jupiters. We also compare stellar metallically, \vsini, flux received by the planet, orbital eccentricity, stellar multiplicity, and spin-orbit misalignment, but find no clear differences between the hot stars that host ``normal" hot Jupiters to those that host the highly inflated hot Jupiters with radii $\geq1.5\rj$. However, we note that several of these host star parameters are poorly constrained for many of the systems. Albedo would be another interesting parameter to compare for these planets but it has not been measured for the majority of planets. It is possible that the highly inflated planets have abnormally low albedos and absorb more energy into their atmospheres at a given stellar insolation. TESS \citep{Ricker14} should provide light curves of sufficient precision to measure secondary eclipses in reflected light for most or all of these systems.

\subsection{Mass-Radius predictions}
Several authors have identified empirical relations that attempt to predict the radius of a planet based on a combination of stellar and planetary properties. \citet{Enoch12} established a relation that predicts planetary radius based only on the equilibrium temperature and orbital semi-major axis. KELT-8 falls into the Jupiter mass bin as defined by \citet{Enoch12} and the relation for that mass regime predicts a radius of 1.43 \rj\ which is far too small relative to our measured value. The most recent empirical relation was presented by \citet{Weiss13}, which uses the incident flux and planetary mass as dependent variables. The \citet{Weiss13} relation predicts a radius of 1.63 \rj\ for KELT-8. If one takes into account the scatter in the \citet{Weiss13} relation (0.1 \rj), and the error on our radius measurement the predicted and measured values agree to within 1.2 $\sigma$.

\subsection{Prospects for atmospheric characterziation}
KELT-8 joins a rare breed of highly inflated hot Jupiters orbiting relatively bright stars. The extended atmosphere, combined with the large transit depth and bright apparent magnitude makes KELT-8 one of the best targets for transmission spectroscopy. In addition, the large planetary radius and high equilibrium temperature give rise to a significant secondary eclipse depth. We calculate an expected eclipse depth of $\approx$1.3 mmag at 3.6 microns by taking the ratio of the blackbody emission from the planet to that of the star and multiplying by $(R_{P}/R_{*})^2$. A secondary eclipse of this depth would be easily accessible by \emph{Spitzer}. Combining $T_{eq}=1675$ from Table \ref{tab:params} with the planetary surface gravity of $\log{g_P}=2.793$, we find that the scale height of a H$_2$-dominated atmosphere ($\mu_m=2$) on KELT-8 would be $H=1113$ km. This large scale height implies that the amplitude of a transmission spectroscopy signal would be $\frac{2R_{P}H}{R_{*}^2}
= 0.22$ mmag \citep[][assuming $H<<R_{P}$]{Winn10}. All of these calculations assume perfect heat redistribution and zero albedo for the planet.  Studies of the atmospheric composition or temperature profile of KELT-8 may provide an explanation for its highly inflated radius.

\subsection{Irradiation History}
We have shown above that KELT-8b is a highly inflated planet, joining the ranks of other hot Jupiters that manifest radii much larger than predicted by standard models for non-irradiated objects with Jovian masses. Several authors \citep[e.g.,][]{Demory11} have suggested an empirical insolation threshold ($\approx 2 \times 10^8$ erg s$^{-1}$ cm$^{-2}$) above which hot Jupiters exhibit increasing amounts of radius inflation. KELT-8b clearly lies above this threshold, with a current estimated insolation of $1.78_{-0.22}^{+0.27}$ \fluxcgs, and therefore its currently large inflated radius is not surprising. At the same time, the KELT-8 host star is found to currently be in a very rapid state of evolution, such that its radius is rapidly expanding as the star crosses the Hertzsprung gap toward the red giant branch. This means that the star's surface is rapidly encroaching on the planet, which presumably is rapidly driving up the planet's insolation and also the rate of any tidal interactions between the planet and the star. 

Therefore it is interesting to consider two questions. First, has KELT-8b's incident radiation from its host star been below the empirical radius inflation threshold in the past? If KELT-8b's insolation only recently exceeded the inflation threshold, the system could then serve as an empirical testbed for the different timescales predicted by different inflation mechanisms \citep[see, e.g.,][]{Assef09,Spiegel12}. Second, what is the expected fate of the KELT-8b planet given the increasingly strong tidal interactions it is experiencing with its encroaching host star? 

\begin{figure}[t] 
   \centering
\includegraphics[width=0.5\textwidth]{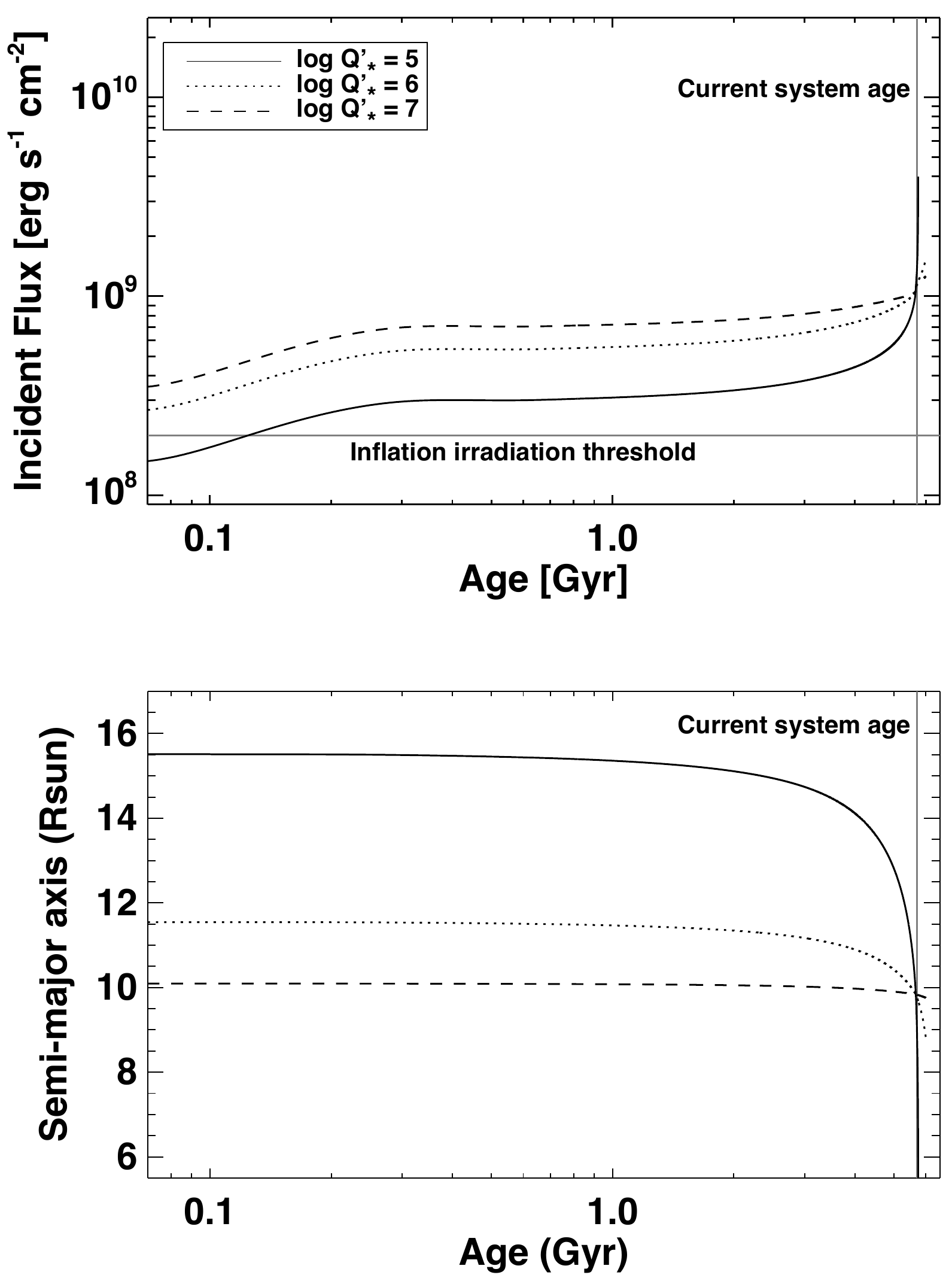}
   \caption{\emph{Top:} Irradiation history of KELT-8b. The insolation received by the planet is well above the empirical inflation irradiation threshold \citep{Demory11} for the entire main-sequence existence of the star except in the case of $\log Q_\star' = 5$ in the very early stages of stellar evolution.
   		\emph{Bottom:} Orbital semi-major axis history of KELT-8b. The planet's semi-major axis is rapidly decreasing as the star evolves off the main sequence. It appears unlikely that KELT-8b will survive past the star's current subgiant phase.}
   \label{fig:irrad}
\end{figure}

To investigate these questions, we follow \citet{Penev14} to simulate the reverse and forward evolution of the star-planet system, using the measured parameters listed in Tables \ref{tab:stellar_params} and \ref{tab:params} as the present-day boundary conditions. This analysis is not intended to examine any type of planet-planet or planet-disk migration effects. Rather, it is a way to investigate (1) the change in insolation of the planet over time due to the changing luminosity of the star and changing star-planet separation, and (2) the change in the planet's orbital semi-major axis due to the changing tidal torque as the star-planet separation changes with the evolving stellar radius. We include the evolution of the star, assumed to follow the Yonsei-Yale stellar model with mass and metallicity as in Table \ref{tab:params}. For simplicity we assume that the stellar rotation is negligible and treat the star as a solid body. We also assume a circular orbit aligned with the stellar equator throughout the full analysis. The results of our simulations are shown in Figure \ref{fig:irrad}. We tested a range of values for the tidal quality factor of the star $Q_\star'$, from $\log Q_\star' = 5$ to $\log Q_\star' = 7$ (assuming a constant phase lag between the tidal bulge and the star-planet direction). $Q_\star'$ is defined as the tidal quality factor divided by the Love number ($Q_\star' = Q_\star/k_2$). We find that although for certain values of $Q_\star'$ the planet has moved substantially closer to its host during the past Gyr, in all cases the planet has always received more than enough flux from its host to keep the planet irradiated beyond the insolation threshold identified by \citet{Demory11}, except perhaps during the pre--main-sequence (prior to an age of $\sim$100 Myr). 

Interestingly, the currently rapid evolution of the star suggests a concomitant rapid in-spiral of the planet over the next few 100 Myr, unless the stellar $Q_\star'$ is large. This planet therefore does not appear destined to survive beyond the star's current subgiant phase. As additional systems like KELT-8b are discovered and their evolution investigated in detail, it will be interesting to examine the statistics of planet survival and to compare these to predictions such as those shown here in Figure \ref{fig:irrad} to constrain mechanisms of planet-star interaction generally and the values of $Q_\star'$ specifically.

\section{Summary}
\label{sec:summary}
We announce the discovery of the highly inflated hot Jupiter, KELT-8b. This planet was initially discovered in KELT photometry, then confirmed via high-precision followup light curves and RVs. We also present adaptive optics imaging of KELT-8 from the Robo-AO system. Astrometry measurements over the past 110 years combined with high resolution spectra of both components firmly establishes the visual companion at 8\farcs8 separation as an unrelated background star. 

We develop a new technique to extract reliable RVs from noisy data that saves a significant amount of telescope time. In the case of RV follow-up for the confirmation or mass measurements of transiting planets where the ephemeris is known and only a few well-timed RV measurements are needed this technique can save 30-50\% of the time by avoiding the need to collect a high quality iodine-free template observation. In addition, lower S/N ratio is required for the RV measurement observations (S/N$\approx$50) through the iodine cell because one component of our model is now completely noise free. Normally, we collect S/N=200 for our precision RV exposures using the standard observed template technique.

KELT-8b has one of the largest radii of any known transiting planet with $R_P = 1.86_{-0.16}^{+0.18}$ \rj. It joins a small and interesting class of highly inflated hot Jupiters orbiting stars slightly more massive than the Sun.

\acknowledgments{
We would like to dedicate this paper to the memory of our partner and collaborator, Giorgio Corfini, who contributed to this discovery and passed away in 2014.
We thank the many observers who contributed to the measurements reported here.
The Robo-AO system was developed by collaborating partner institutions, the California Institute of Technology and the Inter-University Centre for Astronomy and Astrophysics, and with the support of the National Science Foundation under Grant Nos. AST-0906060, AST-0960343 and AST-1207891, the Mt. Cuba Astronomical Foundation and by a gift from Samuel Oschin. Ongoing science operation support of Robo-AO is provided by the California Institute of Technology and the University of Hawai`i. C.B. acknowledges support from the Alfred P. Sloan Foundation.
This material is based upon work supported by the National Science Foundation Graduate Research Fellowship under Grant No. 2014184874. Any opinion, findings, and conclusions or recommendations expressed in this material are those of the authors(s) and do not necessarily reflect the views of the National Science Foundation.
Work by T.G.B., B.S.G., and D.J.S. was partially supported by NSF
CAREER Grant AST-1056524
K.P. acknowledges support from NASA grant NNX13AQ62G.
G.D. acknowledges Regione Campania, POR FSE 2007-13.
This work made use of the SIMBAD database (operated at CDS, Strasbourg, France), 
NASA's Astrophysics Data System Bibliographic Services, 
and the NASA Star and Exoplanet Database (NStED).
This research made use of the Exoplanet Orbit Database
and the Exoplanet Data Explorer at exoplanets.org.
This research has made use of the Washington Double Star Catalog maintained at the U.S. Naval Observatory.
Finally, the authors wish to extend special thanks to those of Hawai`ian ancestry 
on whose sacred mountain of Maunakea we are privileged to be guests.  
Without their generous hospitality, the Keck observations presented herein
would not have been possible.
}

\bibliographystyle{apj}
\bibliography{kelt8}

\enddocument